\documentclass[preprint,12pt]{elsarticle}

\usepackage{amssymb,amsmath,longtable,}
\usepackage{url}
\usepackage{algorithm}
\usepackage{algpseudocode}
\usepackage{xcolor}
\usepackage{colortbl}
\usepackage{xr}
\usepackage{booktabs}

\definecolor{Gray}{gray}{0.85}
\definecolor{LightCyan}{rgb}{0.88,1,1}

\newcolumntype{a}{>{\columncolor{Gray}}c}
\newcolumntype{b}{>{\columncolor{white}}c}

\makeatletter
\newcommand*{\indep}{%
	\mathbin{%
		\mathpalette{\@indep}{}%
	}%
}
\newcommand*{\nindep}{%
	\mathbin{
		\mathpalette{\@indep}{\not}
	}%
}
\newcommand*{\@indep}[2]{%
	\sbox0{$#1\perp\m@th$}
	\sbox2{$#1=$}
	\sbox4{$#1\vcenter{}$}
	\rlap{\copy0}
	\dimen@=\dimexpr\ht2-\ht4-.2pt\relax
	\kern\dimen@
	{#2}%
	\kern\dimen@
	\copy0 
} 
\makeatletter

\newcolumntype{H}{>{\setbox0=\hbox\bgroup}c<{\egroup}@{}}

\begin{document}

\begin{frontmatter}



\title{Unguided structure learning of DAGs for count data}
\author[label1]{Thi Kim Hue NGUYEN}

 \affiliation[label1]{organization={Department of Statistical Sciences,University of Padova},
             addressline={Via C. Battisti, 241},
             city={Padova},
             postcode={35121},
             state={Padova},
             country={Italy}}

\author[label2]{Monica Chiogna}

\affiliation[label2]{organization={Department of Statistical Sciences ``Paolo Fortunati'',University of Bologna},
            addressline={Via Zamboni, 33}, 
            city={Bologna},
            postcode={40126}, 
            state={Bologna},
            country={Italy}}

\author[label1]{Davide Risso}


\begin{abstract}
Mainly motivated by the problem of modelling directional dependence relationships for multivariate count data in high-dimensional settings, we present a new algorithm, called learnDAG, for learning the structure of directed acyclic graphs (DAGs).  In particular, the proposed algorithm tackled the problem of learning DAGs from observational data in two main steps: (i) estimation of candidate parent sets; and (ii) feature selection. We experimentally compare learnDAG to several popular competitors  in recovering the true structure of the graphs in situations where relatively moderate sample sizes are available. Furthermore, to make our algorithm is stronger, a validation of the algorithm is presented through the analysis of real datasets.
\end{abstract}




\end{frontmatter}



\section{Introduction}
\label{sec:intro}

{ 
In various fields such as single-cell sequencing, spatial incidence analysis, and sports science, there has been a noticeable increase in the prevalence of large-scale multivariate count data. Researchers have acknowledged the importance of capturing complex interactions among the variables of interest, leading to the consideration of graphs as modelling tools. Where understanding the direction of association among variables is crucial, like in gene networks,  directed graphs offer distinct advantages. They provide clarity and ease of understanding, catering to researchers with diverse backgrounds. Moreover, those which are acyclic, directed acyclic graphs (DAGs), represent ideal models for estimating causal effects, underscoring the significance of utilizing them to answer various research questions.

In this paper, we tackle the problem of learning the structure of DAGs for such count data. In some circumstances, some ``formal'' or mathematical structure of a system, often abstracting from lower-level details, is available in the form of a topological ordering. Whether or not these topological features are viewed only as explanatory while lacking causal information, this knowledge turns the problem of learning the structure of a DAG into a straightforward task through the strategy of neighbourhood recovery \cite{hue2022guided}. However, in many real situations, such valuable prior information might be unknown or only imprecisely known. For example, when dealing with biological networks known as pathways \citep{kanehisa:2000}, a topological ordering may be available, but it could be misspecified due to an inaccurate representation of the biological system or to the choices made in translating a pathway diagram into a fully DAG. 

}

{

Here, we develop a new algorithm for learning DAGs which does not require prior knowledge of the ordering of variables. In particular, the proposed algorithm consists of three steps: 1) preliminary neighbourhood selection using methods for learning undirected graphs; 2) orienting edges using a log-likelihood score (or a BIC score); and 3) pruning the resulting DAG using variable selection algorithms such as Lasso, or significance tests.

The paper is organized as follows. The proposed algorithm is given in Section  \ref{proposealgorithm}.  Section \ref{Empiricalstudy} provides experimental results, that illustrate the performance of our methods in finite samples.  Biological and sport validation of the algorithm is
presented in Section \ref{Result_realdata}.  Some conclusions and remarks are given in Section \ref{guidremarks}.

\section{The learnDAG algorithm}\label{proposealgorithm}
\noindent
Consider a $p$-dimensional random vector $\mathbf{X}=(X_1,\ldots,X_p)$ such that each random variable $X_s$ corresponds to a node of a directed graph $G=(V,E)$ with index set $V=\{1,2,\ldots,p\}$.  In the Poisson case, the distribution of $\mathbf{X}$ has the form 
\begin{eqnarray}\label{dijoinfull}
\mathbb{P}_{\boldsymbol{\theta}}(\mathbf{x})&=&\! \exp\!\big\{\sum_{j=1}^p\big(\theta_jx_j+\sum_{k\in {pa(j)}}\!\!\!\theta_{jk}x_jx_k-\log x_j!-e^{\!\theta_j+\sum_{k\in {pa(j)}}\!\!\theta_{jk}x_k}\big)\big\}\nonumber\\
&=&\exp\big\{\sum_{j=1}^p\theta_jx_j+\sum_{j=1}^p\sum_{k\ne j}\theta_{jk}x_jx_k-\sum_{j=1}^p\log x_j!-\sum_{j=1}^pe^{\theta_j+\sum_{k\ne j}\theta_{jk}x_k}\big\},\nonumber
\end{eqnarray}
where $pa(j)$ is the set of parents of a vertex
$j$, and  $\theta_{jk}= 0$ if $k\notin pa(j)$. Let $\mathbf{x}^{(1)},\mathbf{x}^{(2)},\ldots,\mathbf{x}^{(n)}$ be $n$ samples independently drawn from the random vector $\mathbf{x}$, with $\mathbf{x}^{(i)}=(x_{i1},x_{i2},\ldots,x_{ip}), ~ i=1,\ldots,n$. Then, the log-likelihood function is of the form:
\begin{eqnarray}\label{dijoinlog}
\ell(\boldsymbol{\theta},\mathbb{X}) &=&\sum_{i=1}^{n}\bigg\{\sum_{j=1}^p\theta_jx_{ij}+\sum_{j=1}^p\sum_{k\ne j}\theta_{jk}x_{ij}x_{ik}-\sum_{j=1}^p\log x_{ij}!-\sum_{j=1}^pe^{\theta_j+\sum_{k\ne j}\theta_{jk}x_{ik}}\bigg\}\nonumber\\
&=& \sum_{j=1}^{p}\ell_j(\boldsymbol{\theta}_{V\backslash\{j\}},\mathbf{x}_{V\backslash \{j\}}),
\end{eqnarray}
where $\ell_j(\boldsymbol{\theta}_{V\backslash\{j\}},\mathbf{x}_{V\backslash \{j\}})=\sum_{i=1}^{n}\big\{(\theta_j+\sum_{k\ne j}\theta_{jk}x_{ik})x_{ij}-\log x_{ij}!-e^{\theta_j+\sum_{k\ne j}\theta_{jk}x_{ik}}\big\}$, denotes the node conditional log-likelihood for $X_j|\mathbf{x}_{V\backslash j}$. Learning DAGs can be performed by estimating parameters $\theta_{jk}$ that maximize the log-likelihood function $\ell(\boldsymbol{\theta},\mathbb{X})$.

{ If the direction of edges is not an object of inference,  several algorithms are available to reconstruct the undirected topology of the graph. See, for example, \cite{JMLR:v22:18-401, yang2012graphical, allen2013local, gallopin2013hierarchical}. However, the inference becomes extremely slippery when adding the need to give also the direction of the edges. In the absence of prior topological knowledge, } the learning process is often divided into two steps: first, the topological ordering is retrieved, and then { directions} are estimated. { In the context of Poisson data, } \cite{park2015learning} proposed to recover the ordering by using an overdispersion score that measures the difference between the conditional mean and variance. However, this approach generally requires a large number of observations and fails with data coming from distributions which are not Poisson. 

Here, we propose a new algorithm, called learnDAG. It consists of three steps: 1) preliminary neighbourhood selection using  methods for learning undirected graphs; 2) orientation of edges using a log-likelihood score (or BIC score); and 3) { possible} pruning of edges for the estimated DAG in 2) using 
significance tests. The pseudo-code of the algorithm for a generic log-likelihood score is presented in Algorithm \ref{learnDAGpseudo}. 
\begin{algorithm}
\small
	\caption{ \label{learnDAGpseudo}learnDAG algorithm. }
	\begin{algorithmic}[1]
		\hrule
		\vskip 2pt
		\State{\textbf{Input}:}  Data containing $n$ independent samples of the $p$-random vector $\mathbf{X}$: $\mathbf{x}^{(1)},\mathbf{x}^{(2)},\ldots,\mathbf{x}^{(n)}$; (and an upper bound \texttt{npa} on the number of parents that a node may have).
		\State{\textbf{Ouput}:} An estimated DAG 
		\State{} \quad  Step 1: Estimate the undirected graph underlying data, construct possible\\ \quad\quad\qquad\quad parent sets $N(j)$.
		\State{} \quad  Step 2:  Orienting edges\\
		 \quad\quad\qquad\quad Perform Algorithm \ref{oriented parent}.\\
		 \quad\quad\qquad\quad Construct potential parent set for each node
		\State{} \quad Step 3: Pruning  the estimated DAG
	\end{algorithmic}
	\hrule
\end{algorithm}

\subsection{Step 1: Preliminary neighbourhood selection (PNS) }
\noindent
{The technique of local search is widely used for learning the structure of a DAG. This method entails the incremental addition of edges and parent sets to nodes. Nevertheless, the scalability of such a technique becomes a challenge as the potential number of parents increases, rendering local search impractical. The main purpose of Step 1 is to possibly overcome such limitations.}

The preliminary neighbourhood selection (PNS)  tries to reduce the cardinality of the candidate set of parents for each node by preliminarly estimating an undirected structure (see Figure \ref{pnsstep}). This step involves utilizing any suitable algorithm for recovering undirected graphs, such as those outlined in \citep{JMLR:v22:18-401, allen2013local, yang2013poisson,yang2015graphical}. For our purposes, we have chosen to utilize a variant of the PC-LPGM algorithm as described in  \cite{JMLR:v22:18-401}. {Here, an edge between nodes $j$ and $k$ is established if the hypothesis of conditional independence  $X_j\indep X_k|\mathbf{X}_{V\backslash \{j,k\}}$ is rejected.} 

{

We implement PNS in a boostrap fashion, using the R function \texttt{boot} from R-package \textit{boot}. Specifically, $B$ random samples of size $n$, $S_1,S_2,\ldots, S_B,$ are generated from the observed data $\mathbf{x}^{(1)},\mathbf{x}^{(2)},\ldots,\mathbf{x}^{(n)}$. For each bootstrap sample $S_i$, an undirected graph is estimated by establishing links between each node $j$ and all other nodes $k$ for which the conditional independence hypothesis is rejected, as outlined in \cite{JMLR:v22:18-401}. The final graph resulting from the bootstrap procedure incorporates only those edges that have been estimated in a minimum of 20\% of the bootstrap samples. Once this estimated undirected graph is obtained, a candidate parent set for each node $j$, denoted as $N(j)$, is straightforwardly constructed, consisting of all nodes connected to node $j$.}

{\color{black} An example of the application of step 1 is given in Figure \ref{pnsstep}. Starting from the fully connected graph on the left hand side, the output of PNS results in the estimated candidate parent sets $N(1)=\{\emptyset\}, N(2)=\{4,5\}, N(3)=\{4,5\}, N(4)=\{2,3,5\}$, and $N(5)=\{2,3,4\}$. }

This simple step considerably reduces the overall computational complexity and running time, especially with large DAGs ($p$ is large), and makes the algorithm feasible up to high-dimensional DAGs.
\begin{figure}[http]
	\begin{center}
		\includegraphics[width = 0.7\linewidth, height=0.25\textheight]{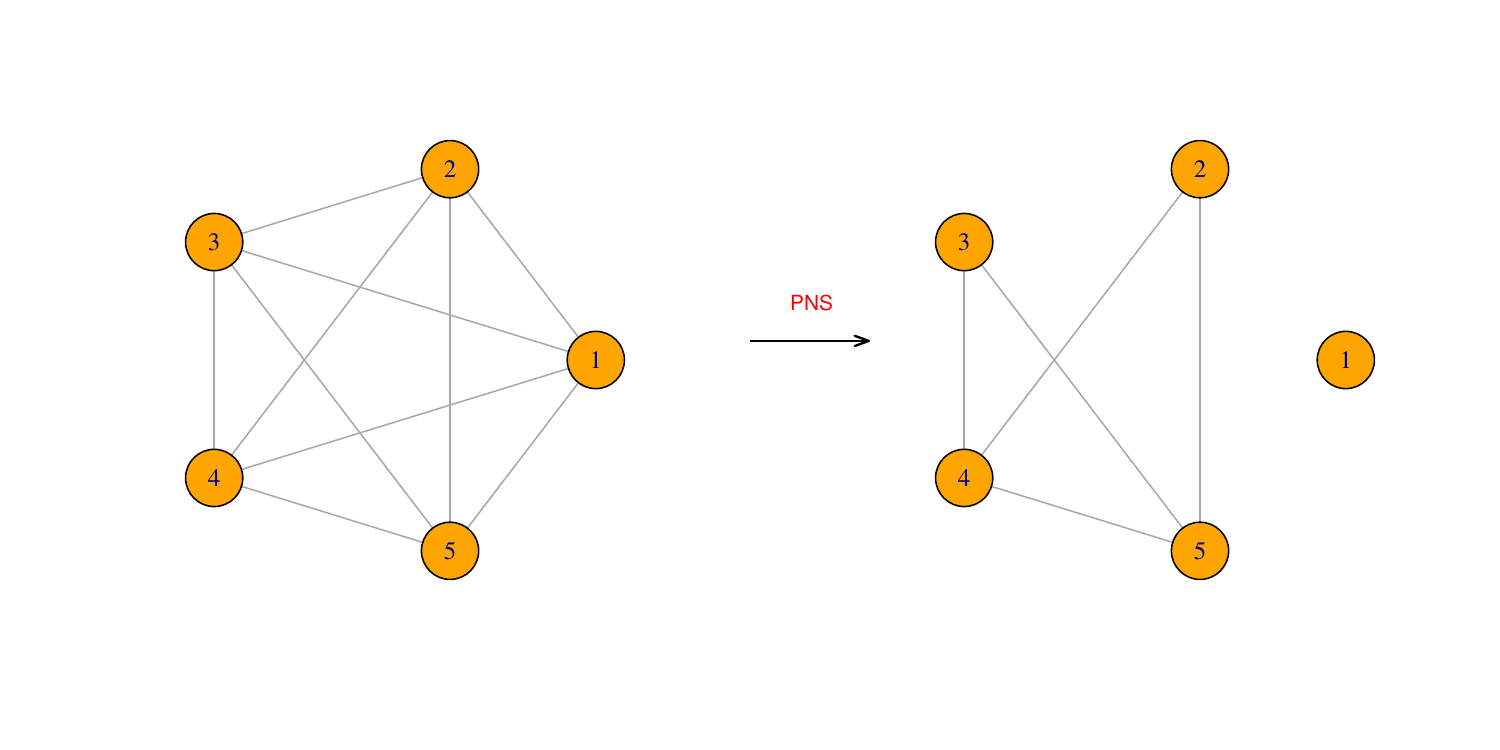}
  \vspace{-1cm}
		\caption{ An example of applying the PNS step on an undirected graph consisting of 6 nodes.}
		\label{pnsstep}
	\end{center}
	
\end{figure}
{ Although particularly recommended for use in high-dimensional regimes, it is however worth mentioning that PNS is not mandatory. If this step is not performed, the candidate set of parents of node $j$ consists of all remaining nodes, i.e.,  $N(j)=V\backslash \{j\},~ j=1,\ldots,p$. }

\subsection{Step 2: Orienting edges }
\noindent
{

Here, we take as input the candidate parent sets $N(j),~ j=1,\ldots,p$, returned from the PNS step, and employ a greedy search to estimate the order of variables. 

{\color{black} In detail, we assume that the distribution of each variable $X_j$, conditional to all possible subsets of variables $\boldsymbol{X}_{\boldsymbol{K}},\, \boldsymbol{K} \subseteq N(j)$ is a Poisson distribution. Then, the algorithm starts from the empty DAG, and at each iteration, the edge $k\rightarrow j$  is added to maximize the  log-likelihood score:
$$\texttt{score}[k,j] = \bigg\{\begin{array}{ccc}
	\ell_j(\hat{\bold{\theta}}_{\hat{pa}(j)\cup \{k\}},\mathbf{x}_{\hat{pa}(j)\cup \{k\}})-\ell_j(\hat{\bold{\theta}}_{\hat{pa}(j)},\mathbf{x}_{\hat{pa}(j)}) &\text{ if }& k\in {N(j),}\\
	-\texttt{Inf} &\text{ if }& k\notin N(j),
	\end{array}$$
where $\ell_j(\cdot)$ denotes the $j$-th conditional log-likelihood in~(\ref{dijoinlog}), and $\hat{pa}(j)$ is an iteratively estimated parent set of the vertex $j$.}

A score matrix, denoted $\texttt{scoremat}$ and containing elements $\texttt{score}[k,j],$ is utilized to track the change in the score function.  
{\color{black}
Figure \ref{step2-learnDAG} shows the application of step 2 to  node 4 of Figure \ref{pnsstep}. The left panel gives an example of the score matrix showing the changing of the log-likelihood score by adding a potential parent ($k\in =N(4)=\{2,3,5\}$) to the current estimate parent set ($\hat{pa}(4)$). The largest gain 0.8 corresponds to the addition of an oriented edge from node $3$ to node $ 4$ (Figure \ref{step2-learnDAG} right).}
\begin{figure}[http]
	\begin{center}
		\begin{minipage}{0.4\linewidth}
			\begin{tabular}{| r r r a r |}
				\toprule
				- & - &-& -& -\\
				- & - & -&0.40&  0.70 \\
				\rowcolor{Gray}
				- & - & -& 0.80& 0.35\\
				- &0.10 &0.50&-& 0.60\\
				- & 0.30& 0.05 & 0.20&- \\
				
				\bottomrule
			\end{tabular}
		\end{minipage}	
		\begin{minipage}{0.4\linewidth}
			\includegraphics[width = 1\linewidth, height=0.2\textheight]{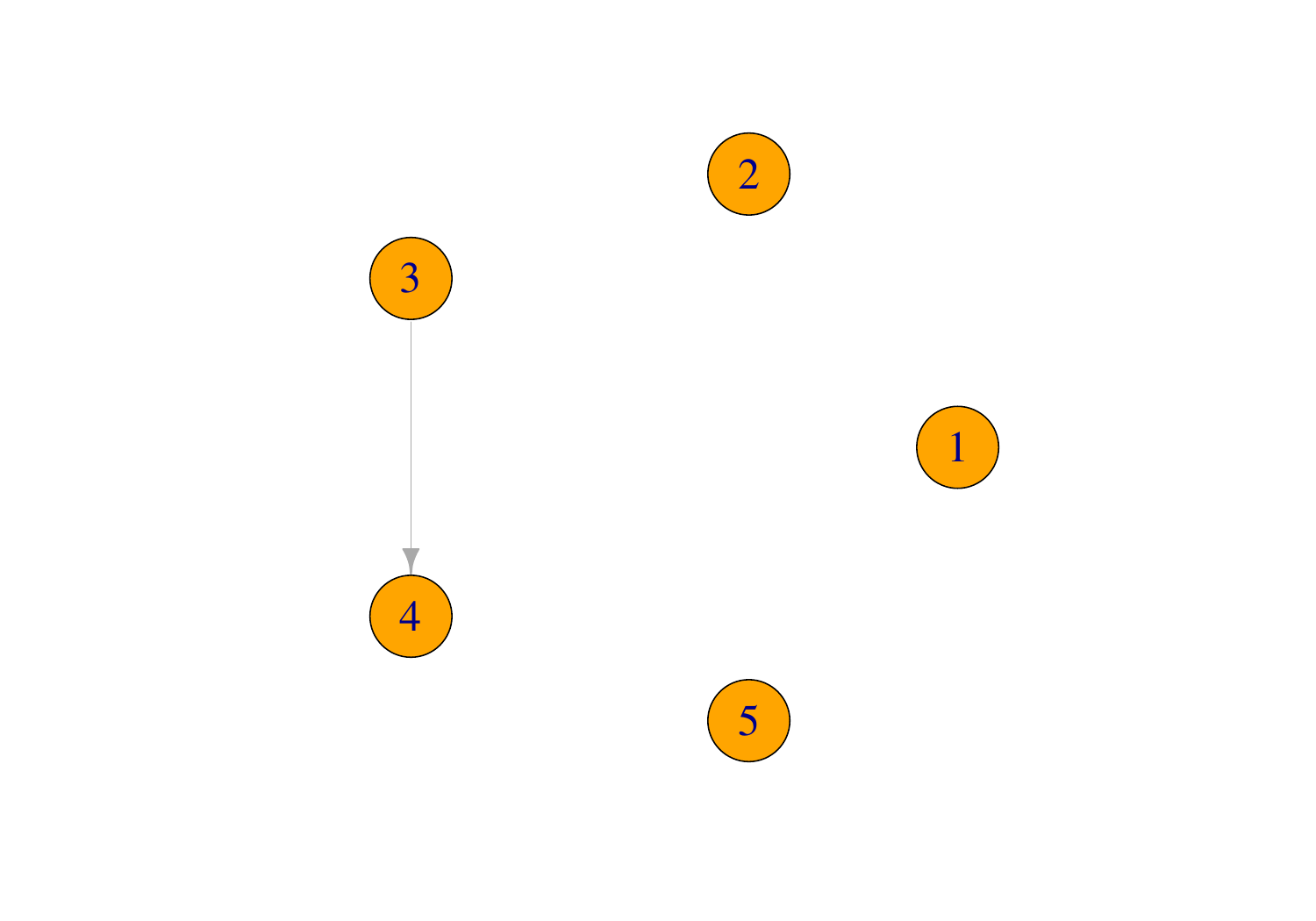}
			
		\end{minipage}
		\caption{An example of adding  edge $3\rightarrow 4$  based on calculating the score matrix.}
		\label{step2-learnDAG}
	\end{center}
\end{figure}

After addition of an edge, the score matrix is updated. Only the $j$-th column needs to be updated,  as the log-likelihood is decomposable into conditional log-likelihoods over all nodes (see expression in~(\ref{dijoinlog})). To avoid cycles, we remove from the matrix all values corresponding to inverse paths on the current graph. The maximum number of iterations is $p(p-1)/2$ corresponding to achieving a fully connected DAG. } 

Once all possible edges have been considered, the potential parent set of each node is defined as the parent set on the resulting DAG. The pseudo-code of this procedure is given in Algorithm \ref{oriented parent}.

\begin{algorithm}
\small
	\caption{\label{oriented parent} orienting edges based on score matrix  algorithm. }
	\begin{algorithmic}[1]
		\hrule
		\vskip 2pt
		\State{\bf Input:}  sets of neighbours $N(j),~ j=1,\ldots,p$. Let $G'$ be an empty DAG on $p$ nodes.
		\State{}Calculate score matrix $\texttt{scoremat}$.
		\State{}   {\bf while} $sum(\texttt{scoremat}~ != -\texttt{Inf}) >0$ {\bf do}
		\State{}   $\quad$ {\bf repeat}
		\State{}   $\quad\quad$ Select the maximum value $\texttt{scoremat}[i,j]$ in $\texttt{scoremat}$.
		\State{}     $\quad\quad$ Add the edge $i\rightarrow j$ to $G'$.
		\State{}     $\quad\quad$ Replace values corresponding to paths and inverse paths on $G'$ by $-\texttt{Inf}$; 
		\State{}     $\quad\quad$ Update $j$-th column of the score matrix.
		\State{}   {\bf until} $sum(\texttt{scoremat}~ != -\texttt{Inf}) =0$.
		\State{\bf return} the potential parent sets obtained from $G'$.
	\end{algorithmic}
	\hrule
\end{algorithm}	


{\color{black}Instead of using the likelihood score as in Algorithm \ref{oriented parent}, other equally valid scoring criteria such as the Akaike Information Criterion and the Bayesian Information Criterion could be adopted.

 It is worth noting that Step 2 allows to further balance the trade-off between goodness of fit and model complexity by adding an upper limit to the number of parents that each node can have, $m$ say. Such value could be determined by some prior knowledge of the sparsity of underlying structures, or, in the absence of any prior knowledge, it can be set to $p-2$.}

If the PNS step is not performed, the above-described procedure can still be performed, but the whole procedure is computationally more expensive, making the algorithm infeasible when the number of variables $p$ is large. 

\subsection{Step 3: Pruning of the DAG}\label{Prun}
\noindent
The purpose of this step is to further refine the structure from estimated graphs after performing (Step 1 and) Step 2. Indeed, 
 depending on the tuning of the previous steps,  findings at this stage of the process might not completely well portray the true structure of the graph. In particular, if loose sparsity conditions are implemented, the graph resulting from Steps 1 and 2 is likely to be a super DAG of the true graph. In these cases, Step 3 aims to enhance the estimation of the graph. 

Available solutions for pruning DAGs could be (i)  {sparse} regression techniques;  and (ii) significance testing procedures. For the first strategy, some methods have been proposed in \cite{friedman2010regularization}.  In detail, for each node $j$, a penalized regression  on the set  representing the potential parents $\hat{pa}(j),~ j=1,\ldots p$, estimated in Step 2. 
Given the solution $\boldsymbol{\hat{\theta}}_{V\backslash\{j\}}$,  the set of parents of node $j$ is given by 
$$\tilde{pa}(j)=\{k\in \hat{pa}(j):\quad \hat{\theta}_{jk}\ne 0\}.$$ 
However, these penalized procedures are scale-variant, a condition that often interferes with some of the filtering steps that are commonly performed { in some domain applications, like in } the analysis of omics data. Moreover,  it could suffer from over-shrinking of small but significant covariate effects.

Here, we consider the second strategy. In particular, we employ hypothesis tests, for example, Wald-type tests on the parameters $\theta_{jk},~k \in \hat{pa}(j).$ We test, at some pre-specified significance level, the null hypothesis $H_0: {\theta}_{jk|\hat{pa}(j)}=0,$  as in \citep{hue2022guided}. If the null hypothesis is not rejected, the edge $k\rightarrow j$ is considered to be absent from the graph. This testing procedure allows to guarantee scale-invariance of the procedure and avoids over-shrinking of small effects. 
 

}


\section{Empirical study}\label{Empiricalstudy}
\noindent
In this section, we present an empirical assessment of our proposed method aimed to ascertain its efficacy in accurately retrieving the true DAG. We also compare our approach with several established competitors to gauge its performance. As measures of ability to recover the true structure of the graphs,  we  adopt three criteria, namely Precision $P$;  Recall $R$; and their harmonic mean, known as $F_1$-score, respectively defined as 
$$ P=\frac{TP}{TP+FP},\, R=\frac{TP}{TP+FN},\, F_1=2  \frac{P .  R}{P+R},$$
where TP (true positive), FP (false positive), and FN (false negative) refer to the { number of} inferred edges \citep{liu2010stability}. 

As competitors, we consider structure learning algorithms for both Poisson and non-Poisson variables. In detail, as representatives of algorithms for Poisson data, we consider: i) the PDN algorithm in \cite{hadiji2015poisson}; ii) the overdispersion scoring (ODS) algorithm in \cite{park2015learning}.  As representatives of algorithms for non-Poisson data, we first consider a structure learning method dealing with the class of categorical data, namely the Max Min Hill Climbing (MMHC) algorithm \citep{tsamardinos2006max}. To apply such algorithms, we categorize our data using the strategy: Gaussian mixture models on log-transformed data shifted by 1 \citep{fraley2002model}. 
Finally, taking into account that structure learning for discrete data is usually performed by employing methods for continuous data after suitable data transformation, we consider one representative of approaches based on the Gaussian assumption,  i.e., the PC algorithm \citep{kalisch2007estimating}, applied to log-transformed data shifted by 1. 


\subsection{Learning algorithms}\label{guidcompals}
\noindent
Acronyms of the considered algorithms are listed below, along with specifications, if needed, of tuning parameters. In this study, we specify the upper limit for the number of parents, $m$, which in this study was set to $m=8$ for $p=10$ and $m=20$ for $p=100$, respectively.

\begin{itemize}

	\item[-]{\bf learnDAG}: the proposed algorithm (Section \ref{proposealgorithm});
	\item[-]{\bf PDN}: Poisson Dependency Networks algorithm  \citep{hadiji2015poisson} with n.trees = 20;
	\item[-]{\bf ODS}:   Overdispersion Scoring algorithm \citep{park2015learning} with $k$-fold cross validation ($k=10$);

	\item[-] {\bf MMHC}:  Max Min Hill Climbing algorithm \citep{tsamardinos2006max} applied to data categorized by mixture models, using $\chi^2$ tests of independence;
	\item[-] {\bf PC}: PC algorithm \citep{kalisch2007estimating}  applied to log-transformed data, using Gaussian conditional independent tests.
\end{itemize}

It is worth noting that the PC algorithm returns partial DAGs (PDAGs) that consist of both directed and undirected edges. In this case, we borrow the idea of  \cite{dor1992simple} to extend a PDAG to DAG. For details of the algorithm, we refer the interested reader to the paper \cite{dor1992simple,hue2022guided}.

\subsection{Results}\label{guidresults}
\noindent
For the two considered vertex cardinalities,  $p=10, 100$;  four chosen sample sizes,  $n=200,500,1000,2000$ for $p=10$, and $n=500,1000,2000,5000$ for $p=100$; and three different structures:
 (i)  scale-free graphs, (ii) hub graphs, (iii) and Erdos-Reny graphs, 50 data were sampled as in \cite{hue2022guided} for each network (see \citep{hue2022guided} for details). 

Table \ref{ttable10-chap2} and Table \ref{ttable100-chap2} report, respectively,  Monte Carlo means of TP, FP, FN, P, R and $F_1$ score for each of the considered method. Each value is computed as an average of the 150 values obtained by simulating 50 samples for each of the three networks. Results disaggregated by network types are given in Table \ref{table10-chap2}, and Table \ref{table100-chap2}.  These results show that our proposed algorithm, learnDAG, is competitive with the other approaches in terms of reconstructing the structure. 

\begin{center}
\fontsize{9.5}{10}\selectfont
		\begin{longtable}{l| l |  r r r r r r r}
			\caption{\label{ttable10-chap2} Monte Carlo marginal means of $TP, FP, FN, P, R$, and $F_1$ score obtained by simulating 50 samples from each of the three networks shown in Figure \ref{DAGtypes10} ($p=10$). The levels of significance of tests $\alpha_{PNS}= \alpha_{Prun}=2(1-\Phi(n^{0.15}))$.}
			\\
			\toprule
			$n$	& Algorithm & $TP$ & $FP$ & $FN$ & $P$ & $R$ &$F_1$  \\
			\midrule
			\endfirsthead
			\multicolumn{8}{c}%
			{{\bfseries \tablename\ \thetable{} -- continued from previous page}} \\
			\toprule
			$n$	& Algorithm & $TP$ & $FP$ & $FN$ & $P$ & $R$ &$F_1$  \\
			\midrule
			\endhead
			
200&learnDAG& 5.613 & 4.973 & 2.720 & 0.536 & 0.670 & 0.591   \\
  &PDN & 6.820 & 28.820 & 1.513 & 0.201 & 0.814 & 0.320 \\ 
  &ODS & 2.667 & 1.236 & 5.681 & 0.714 & 0.315 & 0.422 \\ 
  &PC & 4.062 & 2.000 & 4.283 & 0.654 & 0.484 & 0.550 \\ 
  &MMHC &  2.329 & 3.859 & 6.007 & 0.392 & 0.276 & 0.319 \\ 
  & & & & & & \\
500&learnDAG& 6.253 & 1.593 & 2.080 & 0.796 & 0.748 & 0.768  \\
  &PDN & 7.067 & 22.180 & 1.267 & 0.258 & 0.844 & 0.388 \\ 
  &ODS & 4.347 & 1.813 & 3.987 & 0.714 & 0.520 & 0.593 \\ 
  &PC & 5.450 & 1.839 & 2.886 & 0.746 & 0.654 & 0.694  \\ 
  &MMHC & 3.338 & 4.365 & 5.000 & 0.444 & 0.398 & 0.417 \\
  & & & & & & \\
1000&learnDAG& 7.260 & 1.220 & 1.073 & 0.858 & 0.870 & 0.862 \\
  &PDN & 7.307 & 17.927 & 1.027 & 0.309 & 0.873 & 0.448 \\ 
  &ODS & 5.213 & 2.527 & 3.120 & 0.681 & 0.625 & 0.648 \\ 
  &PC & 6.233 & 1.513 & 2.100 & 0.805 & 0.749 & 0.775 \\ 
  &MMHC & 4.000 & 4.327 & 4.340 & 0.484 & 0.477 & 0.478 \\  
    & & & & & & \\
 2000&learnDAG & 7.780 & 0.933 & 0.553 & 0.896 & 0.933 & 0.913  \\ 
  &PDN & 7.160 & 14.767 & 1.173 & 0.346 & 0.854 & 0.480 \\ 
 &ODS & 6.120 & 2.713 & 2.213 & 0.699 & 0.733 & 0.712 \\ 
   &PC & 6.873 & 1.267 & 1.460 & 0.847 & 0.827 & 0.836 \\ 
  &MMHC & 4.302 & 4.255 & 4.034 & 0.504 & 0.513 & 0.508 \\

			\hline
		\end{longtable}
\end{center}

\begin{center}
\fontsize{9.5}{10}\selectfont
		\begin{longtable}{l| l |  r r r r r r r}
			\caption{\label{ttable100-chap2} Monte Carlo marginal means of $TP, FP, FN, P, R$, and $F_1$ score obtained by simulating 50 samples from each of the three networks shown in Figure \ref{DAGtypes100} ($p=100$). The levels of significance of tests $\alpha_{PNS}= \alpha_{Prun}=2(1-\Phi(n^{0.2}))$.}
			\\
			\toprule
			$n$	& Algorithm & $TP$ & $FP$ & $FN$ & $P$ & $R$ &$F_1$  \\
			\midrule
			\endfirsthead
			\multicolumn{8}{c}%
			{{\bfseries \tablename\ \thetable{} -- continued from previous page}} \\
			\toprule
			$n$	& Algorithm & $TP$ & $FP$ & $FN$ & $P$ & $R$ &$F_1$  \\
			\midrule
			\endhead
			
  500& learnDAG& 48.510 & 11.752 & 52.530 & 0.791 & 0.478 & 0.587\\
  &PDN & 64.773 & 419.773 & 36.227 & 0.216 & 0.637 & 0.250 \\ 
  &ODS & 36.040 & 84.007 & 64.960 & 0.298 & 0.353 & 0.317 \\ 
  &PC & 26.693 & 26.127 & 74.307 & 0.444 & 0.259 & 0.323 \\ 
  &MMHC & 47.993 & 89.340 & 53.007 & 0.348 & 0.474 & 0.401 \\ 
    & & & & & & \\
  1000&learnDAG&71.173 & 17.373 & 29.827 & 0.798 & 0.702 & 0.746 \\
  &PDN & 69.713 & 323.793 & 31.287 & 0.286 & 0.685 & 0.309 \\ 
  &ODS & 45.413 & 83.947 & 55.587 & 0.355 & 0.444 & 0.386 \\ 
  &PC & 34.087 & 32.827 & 66.913 & 0.459 & 0.331 & 0.381 \\ 
  &MMHC & 62.447 & 74.787 & 38.553 & 0.455 & 0.618 & 0.524 \\ 
    & & & & & & \\
  2000&learnDAG&82.893 & 17.780 & 18.107 & 0.821 & 0.819 & 0.820\\
  &PDN & 67.733 & 237.620 & 33.267 & 0.338 & 0.664 & 0.341 \\ 
  &ODS & 54.767 & 82.687 & 46.233 & 0.398 & 0.538 & 0.452 \\ 
  &PC & 41.400 & 39.180 & 59.600 & 0.478 & 0.402 & 0.435 \\ 
  &MMHC & 72.100 & 60.813 & 28.900 & 0.544 & 0.714 & 0.617 \\ 
      & & & & & & \\
  5000&learnDAG&88.893 & 15.753 & 12.107 & 0.848 & 0.879 & 0.863 \\
  &PDN  & 65.020 & 149.351 & 36.007 & 0.406 & 0.636 & 0.380 \\ 
  &ODS & 63.153 & 67.520 & 37.847 & 0.494 & 0.621 & 0.539 \\ 
  &PC& 48.940 & 45.560 & 52.060 & 0.501 & 0.477 & 0.488 \\ 
  &MMHC & 77.248 & 47.168 & 23.544 & 0.621 & 0.766 & 0.685 \\ 
			\hline
			
		\end{longtable}
\end{center}

\begin{figure}[http]
	\begin{center}
		\includegraphics[width = 0.9\linewidth, height=0.42\textheight]{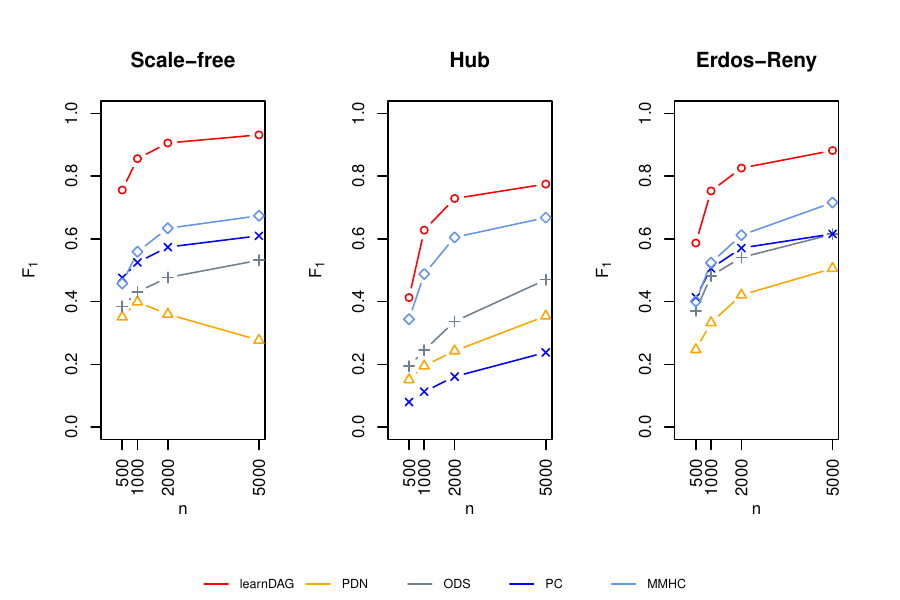}
		\caption{\scriptsize $F_1$-score of the considered algorithms: learnDAG; PDN; ODS; MMHC; PC; for the three types of graphs in Figure \ref{DAGtypes100} with $p=100$ and sample sizes $n=500,1000,2000, 5000.$}
		\label{DAG100F1}
	\end{center}
\end{figure}

When $p=100,$ the algorithm learnDAG reaches the highest $F_1$ score, followed by the ODS, and the MMHC algorithms.  In fact, the $F_1$ score is closed to 1 when $n \ge 5000$ (see Table \ref{ttable100-chap2}, and Figure \ref{DAG100F1}).  This means that the proposed algorithm can recover the underlying DAG using only the given data. 

\begin{figure}[http]
	\begin{center}
		\includegraphics[width = 0.9\linewidth, height=0.69\textheight]{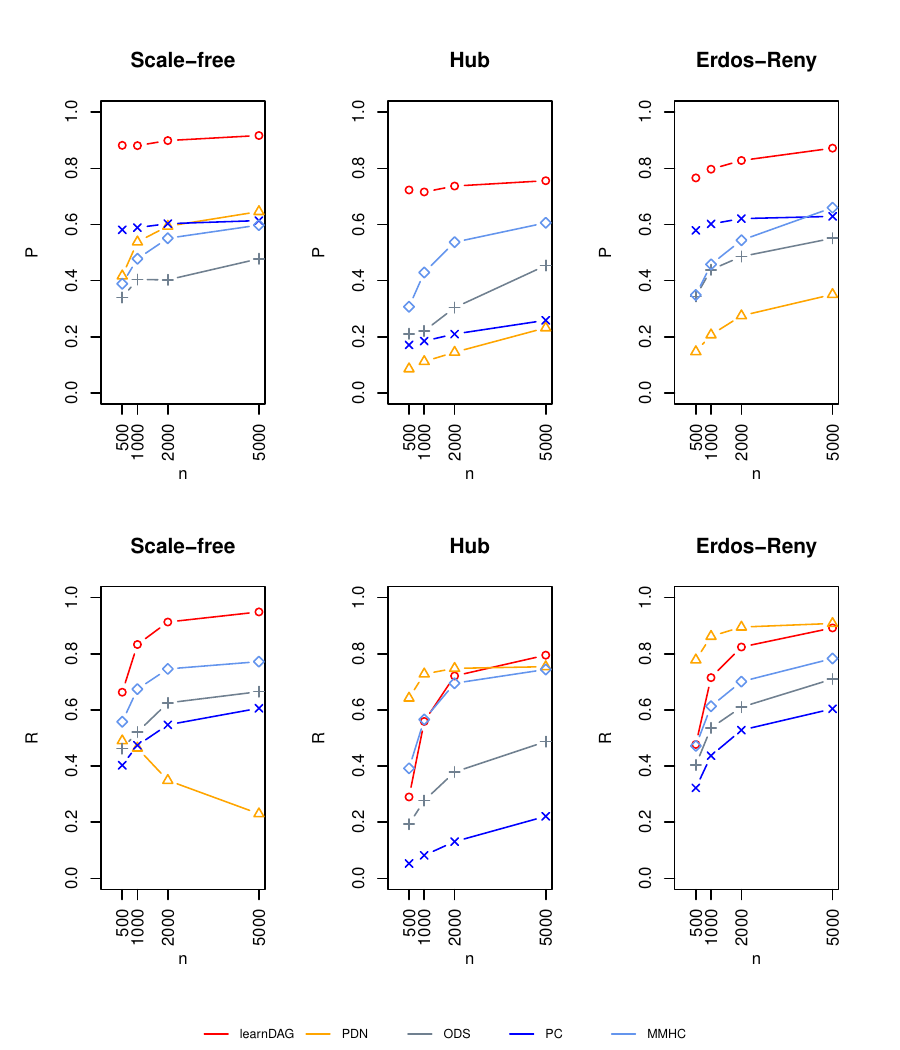}
		\caption{\scriptsize Precision and Recall of the considered algorithms: learnDAG; PDN; ODS; MMHC; PC; for the three types of graphs in Figure \ref{DAGtypes100} with $p=100$ and sample sizes $n=500,1000,2000, 5000.$}
		\label{DAG100PPVSe}
	\end{center}
\end{figure}

 A closer look at the Precision $P$,  and Recall $R$ plots (see Figure \ref{DAG100PPVSe}) provides further insight of the behaviour of considered methods. Among the algorithms with the highest Recall, learnDAG has the highest Precision in all scenarios.

Probably closest in spirit to learnDAG is the ODS algorithm. The performance of the learnDAG algorithm appears to be far better than ODS. Besides using  difference strategies to orient edges, we also need to stress the good performances of learnDAG related to the difference between penalization and testing procedures. This substitution has some advantages over the alternative approach, see \cite{JMLR:v22:18-401} for an extended discussion
of the Poisson case. Moreover, ODS uses the LPGM model \citep{allen2013local} to search the candidate parent sets for each node.  As a consequence, the performance of ODS is highly dependent on the result from LPGM algorithm. However, this result depends on the tuning of its parameters ($\beta$, $\gamma$, $sth$, etc). Here, we used the best combination of parameters that we managed to find in \cite{JMLR:v22:18-401}, i.e., $B=50,$ $nlambda=20$, $\frac{\lambda_{min}}{\lambda_{max}}=0.01$, $\gamma = 10^{-6}$, $sth=0.6$, $\beta=0.1$ for $p=10$ and $\beta=0.05$ for $p=100$.

The performances of  PDN is overall less accurate. This result could come from the problem of inconsistency of PDN in some circumstances and the uncertainties in recovering the direction of interactions.

When considering other methods,  category-based methods (MMHC), and Gaussian-based methods (PC), both perform less accurately as expected. A reason for it is that we are working with a misspecified model, i.e.,  the data generating process is truly Poisson but we transformed data to apply MMHC, and PC algorithm. This approach can work well in some circumstances, however, it could be also ill-suited, for example, the performance of PC with hub graph (see Figure \ref{DAG100F1}).  

We have focused here on $p = 100$, as this setting is closer to our real application, i.e., high dimensional data. Results for the $p=10$ are shown in  Table \ref{ttable10-chap2}, and Appendix, Figure \ref{DAG10F1}, Figure \ref{DAG10PPVSe} lead to similar conclusions.

\section{Results on real data}\label{Result_realdata}
{In this section, we showcase our method on two real datasets from different fields.  First, we focus on the single-cell sequencing data from \cite{brann2020non}, with the aim of inferring  the activation order of the genes in the Wnt singalling pathway. We then turn our attention to sport data science, reanalyzing the NBA Player Statistics dataset presented in \citep{park2018learning}. Here, the aim is to evaluate the ability of our proposal to recover logical relationships from the data (e.g., shooting should precede scoring and not viceversa) and compare its performance to other DAG learning approaches.}

\subsection{Preprocessing}\label{preprocess}

As measurements were zero-inflated and highly skewed, standard preprocessing was applied to the data  as in \cite{ JMLR:v22:18-401,allen2013local}. 
In particular, we first normalized the data by $95\%$ quantile matching between statistical units to account for differences in sequencing depths in single cell RNA sequencing data, and {player position characteristics} in NBA data. Second, we adjust the data to be closer to a Poisson distribution by using a power transformation $X^\alpha$,  where $\alpha \in (0,1]$ is chosen to minimize the distance between the empirical distribution and the Poisson distribution, measured by Kolmogorov-Smirnov statistics. Finally, we round the normalized data to the closest smaller integer (flooring). See \citep{JMLR:v22:18-401, allen2013local, nguyen2023structure} for discussions on the need of using  preprocessing steps.

\subsection{Wnt signalling pathway}
Our first analysis focuses on  a set of cells, assayed with 10X Genomics (v2 chemistry) after injury of the mouse olfactory epithelium (OE), to characterise {the regeneration of olfactory neuons by stem cells following injury \citep{brann2020non}.  After following the data preprocessing as in \cite{brann2020non}, we obtain a dataset consisting of 7782 stem cells (also known as Horizontal Basal Cells, denoted by HBCs), 5418 activated stem cells (activated HBCs, denoted by HBC*), 755 Globose Basal Cells (GBCs), 2859 immature olfactory neurons (iOSN), and 929 mature olfactory neurons (mOSN).}

\begin{figure}[http]
	\begin{center}
  \includegraphics[width = 1\linewidth, height=0.432\textheight]{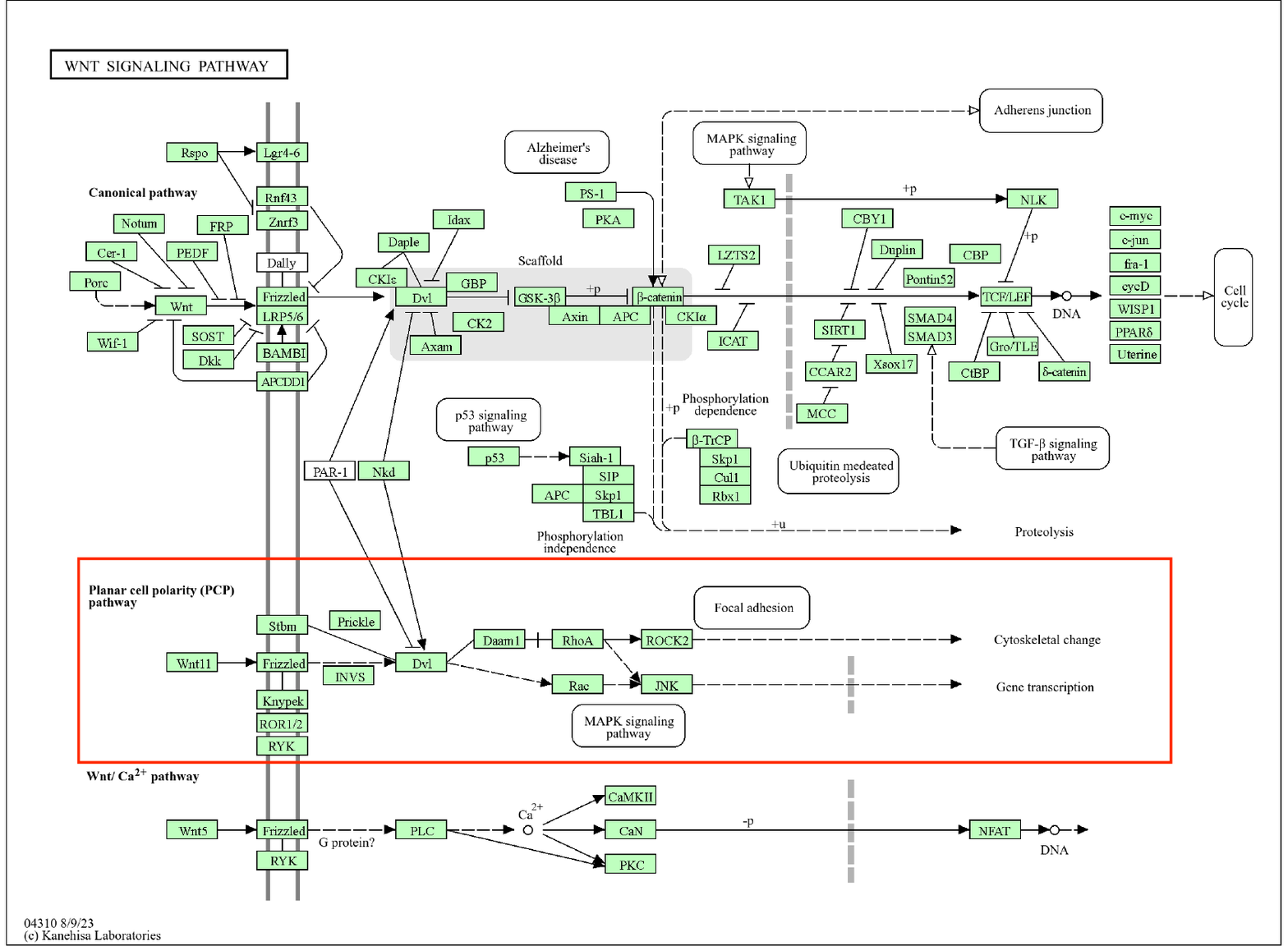}
		\caption{  Wnt signaling pathway- Mus musculus (house mouse) from KEGG. Planar cell polarity (PCP) pathway is highlighted in red box.}
		\label{wnt_pathway}
	\end{center}
\end{figure}

We analyzed a set of 171 genes, annotated with the term ``Wnt signalling pathway'' in the KEGG database \citep{kanehisa:2000} (see Figure \ref{wnt_pathway}), expressed in the activated HBCs. After performing preprocessing steps as described in Section \ref{preprocess}, we obtain a dataset consisting of 5356 cells and 123 genes.

Our goal here is to infer the topological ordering of genes, with a particular focus on the three different Wnt pathways: the canonical pathway, the planar cell polarity (PCP) pathway and the Wnt/Ca2+ pathway.  
We expect to recover the order activation of genes in the Wnt signalling pathway.%

\begin{figure}[http]
	\begin{center}
		\includegraphics[width = 1\linewidth, height=0.47\textheight]{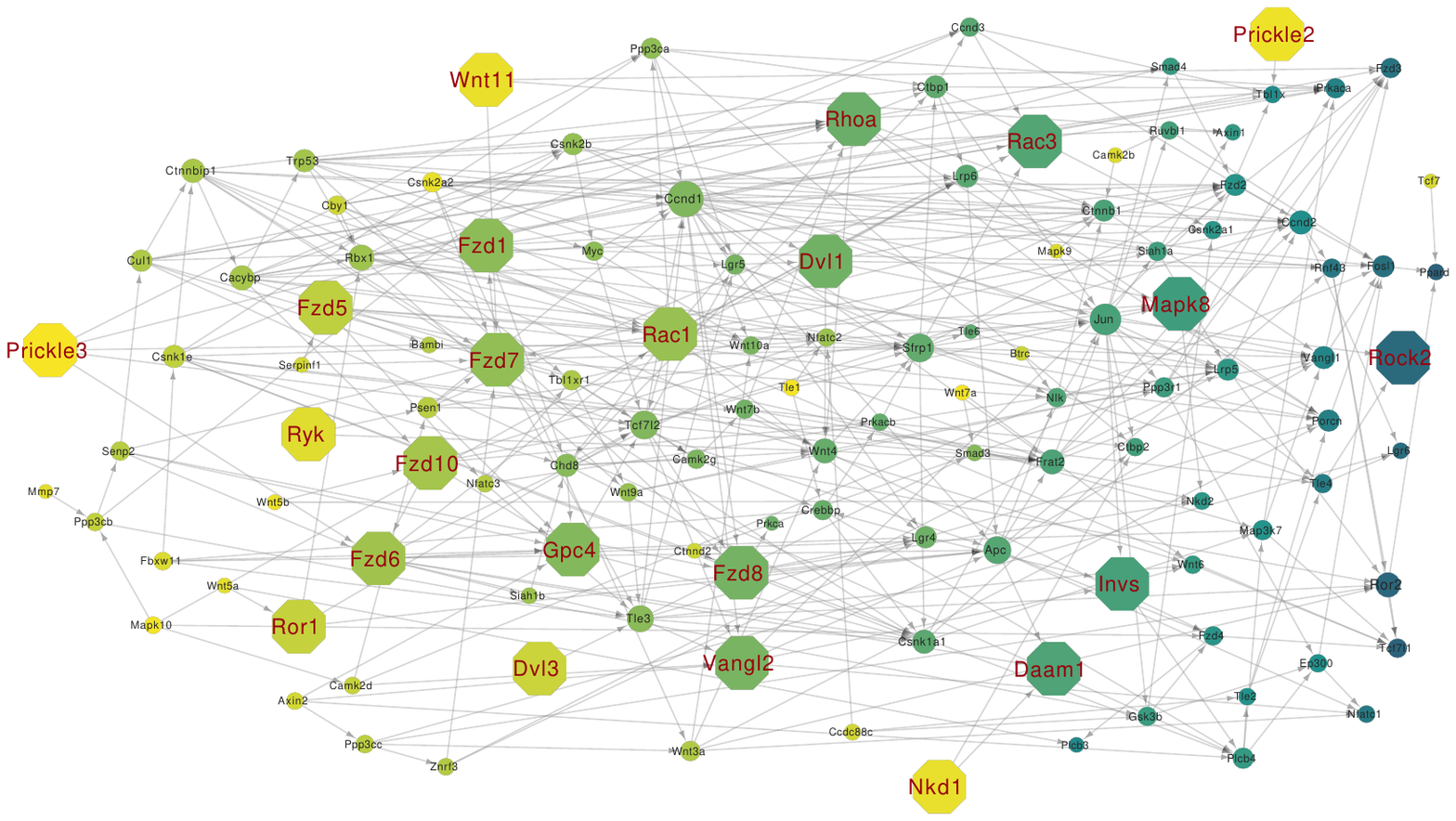}
  \vspace{-2.5cm}
 
		\caption{  Network of Wnt signalling pathway DAG estimated with learnDAG. Planar cell polarity (PCP) signalling genes are displayed with an Octagon, and the remaining nodes are displayed with a circle. The gene colour (from yellow to dark green) is proportional to a topological ordering of the DAG estimated by our approach.}
		\label{OEresult}
	\end{center}
\end{figure}

Normalized data were used as input to learnDAG. Turning parameters $\alpha_{PNS}= \alpha_{Prun}=2(1-\Phi(n^{0.15}))$, and $m=8$ resulted in
a spare graph, shown in Figure \ref{OEresult}.

To better visualise the result, we focus on the genes that belong to the PCP pathway (displayed with an Octagon, and bigger size). By comparing our estimated DAG with the PCP pathway (Figure \ref{wnt_pathway}) from the Wnt signalling pathway (\url{https://www.genome.jp/pathway/mmu04310}), Figure \ref{OEresult} shows that our algorithm can reconstruct, from gene expression data, a biologically meaningful structure that confirms biological processes. 
For instance, Figure \ref{OEresult} shows experimental evidence that Frizzled (Fzd), Prickle, and Stbm (Vangl2) influence the activation of Rho GTPases and JNK(Mapk) through Dvl and Dishevelled-associated activator of morphogenesis 1 (Daam1) \citep{yang2015wnt}. This is coherent with the fact that Vangl2, the Fzd and Prickle gene families (Fzd1, Fzd5, Fzd6, Fzd7, Fzd8, Fzd10, Prickle2, Prickle3) appear before Dvl, Daam1 and Mapk in the topological ordering of our estimated DAG, as indicated in Figure \ref{OEresult} by the lighter color.
Moreover, in the PCP signaling, Wnt binds to the receptor complex and 
lead to the activation of the small GTPases Rhoa (Ras homologue gene-family member A) and Rac \citep{mezzacappa2012activation}. This is consistent with Rhoa and Rac being descendants of Wnt11 in the topological ordering. Activation of Rho GTPase leads to the activation of the Rho-associated kinase (ROCK), which leads to remodelling of the cytoskeleton \citep{habas2003coactivation}. This is coherent with {\color{black} Rock2 appears as the final node of the PCP pathway. }
The fact that our algorithm reconstructs these known topological ordering holds promise for reconstructing the underlying structure, i.e., gene interactions.

\subsection{NBA Player Statistics}
Our second analysis focuses on a set of 441 NBA player statistics during the 2009/2010 season (see R package SportsAnalytics for details, \url{https://cran.r-project.org/src/contrib/Archive/SportsAnalytics/}). {The data recorded 24 variables of 441 players including player name, team name, players position and a set of statistics on their performance in the games. We followed the same data filtering as in \citep{park2018learning}, and the data preprocessing as in Section \ref{preprocess}, which resulted in a dataset containing 18 variables and 441 players.} 

Our goal here is to compare the performance of the proposed method to its competitors on real data. In detail, we compare our proposal to the closet algorithm in spirit, i.e.,  the ODS algorithm. For a fair comparison, tuning parameters are chosen such that the numbers of edges in the resulting graphs are comparable to the ones recorded by applying the ODS algorithm in previous studies.

\begin{figure}[http]
	\begin{center}
		\includegraphics[width = 1\linewidth, height=0.3\textheight]{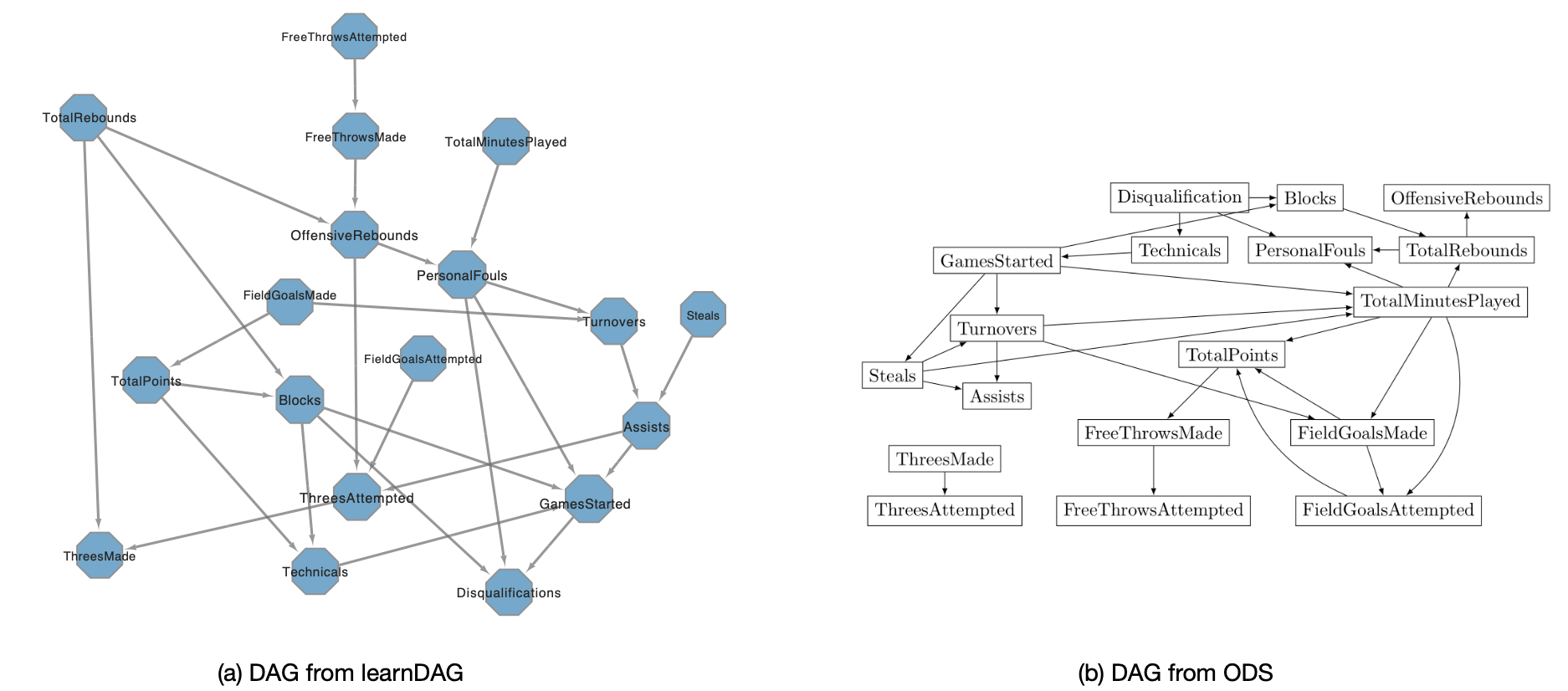}
		\caption{  Networks of NBA players statistics estimated by  learnDAG (a), and ODS (b) for Poisson DAG models. The tuning parameters $\alpha_{PNS}= \alpha_{Prun}=0.05$}
		\label{NBA_large}
	\end{center}
\end{figure}

The learnDAG algorithm inferred a sparse graph with tuning parameters $\alpha_{PNS}= \alpha_{Prun}=0.05$, and $m=10$, shown in Figure \ref{NBA_large}(a). It is immediate to recognize some {logically consistent directional relationships between variables}. For instance, the number of Free Throws Attempted implies the number of Free Throws Made, the number of Three Attempted
implies the number of Three Made, and the number of Total Rebounds implies the number of Offensive Rebounds. Compared to the graph obtained from the ODS algorithm, {reproduced in Figure \ref{NBA_large}(b)}, we see that while the two resulting graphs share some common edges, such as Total Minutes Played $\rightarrow$ Personal Fouls, Field goals made $\rightarrow$ total points, {the direction of learnDAG's inferred edges are generally more meaningful than those of ODS. For instance, learnDAG correctly infers Three Attempted $\rightarrow$ Three Made, Free Throws Attempted $\rightarrow$ Free Throws Made, and Personal Fouls $\rightarrow$ Disqualification. Conversely, while ODS inferred the same edges, their direction was reversed.} 

\begin{figure}[http]
	\begin{center}
		\includegraphics[width = 1\linewidth, height=0.28\textheight]{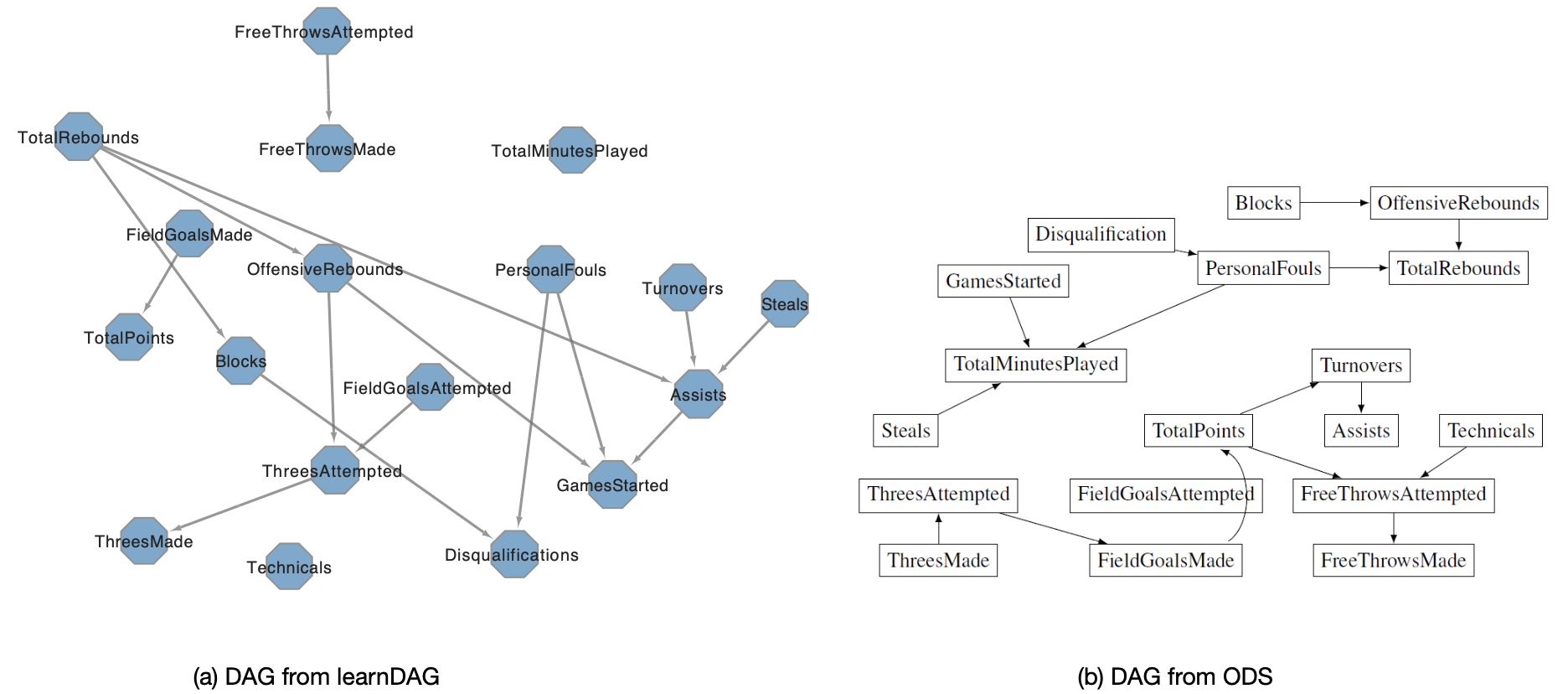}
		\caption{  Networks of NBA players statistics estimated by  learnDAG (a), and ODS (b) for Poisson DAG models. The tuning parameters $\alpha_{PNS}= \alpha_{Prun}=2(1-\Phi(n^{0.15}))$.}
		\label{NBA_small}
	\end{center}
\end{figure}

Figure \ref{NBA_small}(a) shows the estimated directed graph using smaller tuning parameters $\alpha_{PNS}= \alpha_{Prun}=2(1-\Phi(n^{0.15}))$. As expected, the estimated DAG has fewer edges than  Figure \ref{NBA_large}(a). {However, it is not always easy to evaluate whether the removed edges have been correctly trimmed: for instance, the estimated DAG in Figure \ref{NBA_small}(a) excludes  edges such as Assist $\rightarrow$ Threes Attempted or Field Goal Made $\rightarrow$ Turnovers, that on the surface may appear unrealistic, but that could be justified by latent characteristics of the players, e.g., players that score more may attempt more difficult plays, which result in turnovers. Overall, the usual trade-off between sensitivity and specifity applies. In any case, compared to the result obtained from ODS with large tuning parameters \citep{park2019identifiability}, reproduced in Figure \ref{NBA_small}(b), learnDAG provides more legitimate directed edges than the ODS algorithm even in the case of increased sparsity. The better performance of learnDAG could be due to the use of testing procedures in Step 1, which reduces the search space of DAGs better than a lasso procedure based on cross validation (see Section \ref{guidremarks} for details).}

{We acknowledge that our estimated DAGs may suffer from the presence of confounding variables not included in the model. Specifically, the players position may impact their statistics, e.g., a point guard is more likely to score threes and to make assists. As a consequence, the directed graph for each position may not be the same. However, the data consists of only 441 players, and does not allow us to learn the underlying graphical structure for each position separately. Nonetheless,  our analyses show that learnDAG leads to a more interpretable graph  than ODS.}


\section{Discussion and conclusions}\label{guidremarks}
\noindent
{\color{black}We have proposed an unguided structure learning algorithm for structure learing of DAGs and compared it to several different approaches. A fundamental aspect of our methodology involves separating identification of potential parent sets from edge selection. This makes the algorithm flexible and easy to implement. On the synthetic datasets considered in Section \ref{Empiricalstudy}, we showed that the algorithm outperforms its natural competitors in terms of reconstructing the structure from given data for
large enough sample sizes, while providing  coherent information and insight on
the real datasets analyzed in Section \ref{Result_realdata}.} 

The final performance of learnDAG { in finite samples} depends on the chosen combination and calibration of three steps. 

\noindent
{
In the context of Step 1, any structure learning algorithm designed for undirected graphs, including but not limited to LPGM \cite{allen2013local} and PC-LPGM \cite{JMLR:v22:18-401}, can be employed. It is evident that each option has its own set of advantages and constraints. As a result, no single option can claim comprehensive superiority. We have a preference for algorithms that are grounded in testing procedures, such as PC-LPGM, which have demonstrated superior performance compared to state-of-the-art algorithms (refer to [3] for further details).

\begin{table}[ht]
\fontsize{7.2}{9}\selectfont
\centering
\caption{\label{table-margin}{ Monte Carlo  means of Precision (P), Recall (R), $F_1$-score, and runtime obtained by simulating 50 replicates of the DAGs shown in Figure \ref{DAGtypes10} in of the main paper  for $p= 10$ variables with Poisson node conditional distribution.  The levels of significance of tests $\alpha_{margin}=0.05,\, \alpha_{PNS}=\alpha_{Prun}=2(1-\Phi(n^{0.15}))$ for $p=10$, and $\alpha_{margin}=0.05,\,\alpha_{PNS}=\alpha_{Prun}=2(1-\Phi(n^{0.2}))$ for $p=100$.}}
\vspace{0.2cm}
\begin{tabular}{r|r|rrrr|rrrr}
  \hline
  &&\multicolumn{4}{c|} {\bf learnDAG}&\multicolumn{4}{c} {\bf learnDAG.margin}\\
Graph&$n$& P & R&$F_1$ & time & P & R&$F_1$ & time   \\ 
    \hline
scalefree&200    &  0.550 &  0.720 &  0.620 &  7.600 &  0.630 &  0.690 &  0.660 &  7.680 \\ 
  &500 &  0.800 &  0.810 &  0.800 & 10.760 &  0.810 &  0.810 &  0.810 & 10.750 \\ 
&1000 &  0.920 &  0.930 &  0.930 & 11.700 &  0.920 &  0.940 &  0.930 & 11.470 \\ 
  &2000 &  0.960 &  0.960 &  0.960 & 11.100 &  0.960 &  0.960 &  0.960 & 10.570 \\ 
    &&&&&&&&\\
 Hub&200 &  0.540 &  0.750 &  0.630 &  8.880 &  0.650 &  0.700 &  0.670 &  9.120 \\ 
  &500 &  0.880 &  0.820 &  0.850 &  9.450 &  0.890 &  0.820 &  0.850 &  9.480 \\ 
 &1000 &  0.920 &  0.930 &  0.920 &  9.680 &  0.920 &  0.930 &  0.920 &  9.530 \\ 
  &2000 &  0.960 &  0.970 &  0.970 & 10.320 &  0.960 &  0.970 &  0.970 &  9.740 \\ 
    &&&&&&&&\\
  Random&200 &  0.420 &  0.490 &  0.450 &  7.620 &  0.500 &  0.470 &  0.480 &  7.580 \\ 
  &500 &  0.760 &  0.650 &  0.690 &  8.190 &  0.760 &  0.640 &  0.690 &  7.910 \\ 
  &1000 &  0.780 &  0.710 &  0.740 &  8.920 &  0.770 &  0.710 &  0.740 &  8.430 \\ 
  &2000 &  0.890 &  0.890 &  0.890 & 11.540 &  0.890 &  0.880 &  0.890 & 10.710 \\

  \hline
\end{tabular}
\end{table}

Moreover, it is important to recognize that in certain cases, such as with small graphs (e.g., when $p$ is small), Step 1 may be unnecessary. From Table \ref{table10-noPNS}, it is evident that for small graphs, the learnDAG algorithm performs similarly with or without Step 1, but the execution time of learnDAG with Step 1 is approximately ten times longer than without. However, implementation of Step 1 is crucial to ensure the practicality of the proposed algorithm in situations with a high dimensionality (i.e., when $p$ is large). For example, when $p=100$, running the learnDAG algorithm without Step 1 takes approximately five times longer to generate results compared to when Step 1 is included (see Table \ref{table10-noPNS} for more information). Thus, it is advisable for users to integrate Step 1 when working with high-dimensional data  and to omit this step for low-dimensional data.

\begin{table}[ht]
\fontsize{7.2}{9}\selectfont
\centering
\caption{\label{table10-noPNS}{ Monte Carlo  means of Precision (P), Recall (R), $F_1$-score, and runtime obtained by simulating 50 replicates of the DAGs shown in Figure \ref{DAGtypes10}, \ref{DAGtypes100} in of the main paper  for $p= 10,100$ variables with Poisson node conditional distribution.  The levels of significance of tests $\alpha_{PNS}=\alpha_{Prun}=2(1-\Phi(n^{0.15}))$ for $p=10$, and $\alpha_{PNS}=\alpha_{Prun}=2(1-\Phi(n^{0.2}))$ for $p=100$.}}
\vspace{0.2cm}
\begin{tabular}{r|r|rrrr|rrrr}
  \hline
  &&\multicolumn{4}{c|} {\bf learnDAG.test}&\multicolumn{4}{c} {\bf learnDAG.noPNS}\\
$p$&$n$& P & R&$F_1$ & time & P & R&$F_1$ & time   \\ 
    \hline
10&500   &  0.680 &  0.610 &  0.640 & 10.350 &  0.740 &  0.640 &  0.680 &  0.740 \\ 
&1000  &  0.810 &  0.750 &  0.770 & 11.030 &  0.780 &  0.750 &  0.760 &  0.940 \\ 
&2000  &  0.890 &  0.880 &  0.880 & 12.200 &  0.860 &  0.890 &  0.880 &  1.360 \\ 

&&&&&&&&\\
100&500  &      0.770 &   0.490 &   0.600 &  49.450 &   0.710&   0.519  & 0.599 & 253.585\\
&1000 &     0.800 &   0.720 &   0.750 &  82.010 &0.771 &  0.705 &  0.736& 396.756\\ 
&2000 &    0.830 &   0.820 &   0.830 & 150.960&0.816 &  0.823  & 0.819 & 679.547 \\ 
&5000  &     0.870 &   0.890 &   0.880 & 399.230 &0.864  &  0.887&    0.876&  1557.897\\ 

  \hline
\end{tabular}
\end{table}

In the context of Step 2, the utilization of the log-likelihood score as the primary tool for orienting the directions of the edges enhances the generalisability of our algorithm to other parametric assumptions compared to methods relying on overdispersion scoring, as seen in \cite{park2015learning}, which are tailored specifically for Poisson data. Furthermore, our approach allows for the use of other likelihood-based scores such as BIC or AIC, thus enabling a broader range of research goals to be considered. Table \ref{table10-step2} presents the Monte Carlo means of Precision (P), Recall (R), F1-score, and runtime of learnDAG when employing different scores to orient the direction of edges. It is evident that learnDAG.loglik and learnDAG.BIC yield similar performance, while learnDAG utilizing the same orientation strategy as the PC algorithm, i.e., learnDAG.oriented, exhibits lower accuracy with a reduced $F_1$ score. This distinction becomes particularly pronounced when analyzing graphs with high dimensions ($p = 100$).

\begin{table}[ht]
\fontsize{6}{9}\selectfont
\centering
\caption{\label{table10-step2}{ Monte Carlo means of Precision (P), Recall (R), $F_1$-score, and runtime obtained by simulating 50 replicates of the DAGs shown in Figure \ref{DAGtypes10}, \ref{DAGtypes100} in of the main paper for $p= 10,100$ variables with Poisson node conditional distribution. Three algorithms are considered: (i) learnDAG using the log-likelihood score,  learnDAG.loglik; (ii) learnDAG using the BIC score, learnDAG.BIC; and (iii) learnDAG using the same strategy as in the PC algorithm, learnDAG.oriented.  The levels of significance of tests $\alpha_{PNS}=\alpha_{Prun}=2(1-\Phi(n^{0.15}))$ for $p=10$, and $\alpha_{PNS}=\alpha_{Prun}=2(1-\Phi(n^{0.2}))$ for $p=100$.}}
\vspace{0.2cm}
\begin{tabular}{r|r|rrrr|rrrr|rrrr}
  \hline
  &&\multicolumn{4}{c|} {\bf learnDAG.oriented}&\multicolumn{4}{c|} {\bf learnDAG.BIC}&\multicolumn{4}{c} {\bf learnDAG.loglik}\\
$p$&$n$& P & R&$F_1$ & time & P & R&$F_1$ & time & P & R&$F_1$ & time  \\ 
 \hline
10&500 &   0.740 &   0.600 &   0.660 &   1.120 &   0.700 &   0.600 &   0.640 &  11.100 &   0.700 &   0.600 &   0.640 &  11.100 \\ 
&1000 &   0.770 &   0.700 &   0.730 &   1.050 &   0.800 &   0.740 &   0.770 &  11.190 &   0.800 &   0.740 &   0.770 &  11.190 \\ 
&2000 &   0.790 &   0.780 &   0.780 &   1.160 &   0.890 &   0.880 &   0.890 &  12.100 &   0.890 &   0.880 &   0.890 &  12.100 \\ 
&&&&&&&&&&&&&\\
100&500 &   0.610 &   0.370 &   0.460 &   4.570 &   0.770 &   0.480 &   0.590 &  49.250 &   0.770 &   0.480 &   0.590 &  49.210 \\ 
&1000 &   0.620 &   0.530 &   0.570 &   6.310 &   0.800 &   0.710 &   0.750 &  81.930 &   0.800 &   0.710 &   0.750 &  81.910 \\ 
&2000&   0.620 &   0.600 &   0.610 &   9.230 &   0.830 &   0.820 &   0.830 & 150.210 &   0.830 &   0.820 &   0.830 & 150.200 \\ 
&5000&   0.630 &   0.630 &   0.630 &  18.110 &   0.870 &   0.890 &   0.880 & 394.510 &   0.870 &   0.890 &   0.880 & 394.530 \\ 
  \hline
\end{tabular}
\end{table}

{\color{black} In Step 3, the objective is  to further refine the structure from estimated graphs  by (Step 1 and) Step 2. It could indeed be the case that is the estimated DAG from the previous steps is a super DAG of the true DAG. This refinement can be achieved through pruning techniques such as sparse regression or significance testing procedures. In this context, we used  significance testing procedures since they offer distinct advantages, as discussed in Section \ref{Prun}.   Any consistent conditional independence tests can be applied in Step 3. However, the convergence rate as well as the properties of the tests will be different. 

It is worth to acknowledge that when the log-likelihood score is employed to direct edges, hypothesis testing can be directly integrated into Step 2.  Specifically,  a deviance test statistic for testing absence of the edge $k\rightarrow j$, i.e., the hypothesis $H_0: \theta_{jk|pa(j)}= 0$  can be naturally obtained as
\begin{eqnarray*}
   D(jk|\hat{pa}(j))&=&2\times \texttt{score}[k,j] \\
   &=&2\times\big(\ell_j(\hat{\bold{\theta}}_{\hat{pa}(j)\cup \{k\}},\mathbf{x}_{\hat{pa}(j)\cup \{k\}})-\ell_j(\hat{\bold{\theta}}_{\hat{pa}(j)},\mathbf{x}_{\hat{pa}(j)})\big).
\end{eqnarray*}
Subject to appropriate regularity conditions,  $D(jk|\hat{pa}(j))$ is asymptotically chi-squared distributed with 1-degree of freedom under the null hypothesis. Hence, row 6 in Algorithm 2 could be substituted with: ``Add the edge $i \rightarrow j$ to $G'$ if $H_0$ is rejected". 
Table \ref{table-noPrun} presents the Monte Carlo means of Precision (P), Recall (R), F1-score, and runtime of learnDAG with Step 3, and learnDAG with deviance tests, called learnDAG.noPrun. The table shows that  learnDAG  and learnDAG.noPrun yield almost identical performance for both low dimension ($p=10$), and high dimensions ($p = 100$). However, in the realm of algorithm design, a hybrid approach offers the flexibility to combine strategies in a versatile manner at each step. This allows for a more dynamic and adaptable decision-making process compared to algorithms that rigidly bind each step to specific choices. It is our opinioin that 
the 3-step approach offers the advantage of being able to tailor the algorithm to the specific requirements of a given problem, ultimately leading to superior performance and outcomes.

\begin{table}[ht]
\fontsize{7.2}{9}\selectfont
\centering
\caption{\label{table-noPrun}{ Monte Carlo  means of Precision (P), Recall (R), $F_1$-score, and runtime obtained by simulating 50 replicates of the DAGs shown in Figure \ref{DAGtypes10}, \ref{DAGtypes100} in of the main paper  for $p= 10,100$ variables with Poisson node conditional distribution.  The levels of significance of tests $\alpha_{PNS}=\alpha_{Prun}=2(1-\Phi(n^{0.15}))$ for $p=10$, and $\alpha_{PNS}=\alpha_{Prun}=2(1-\Phi(n^{0.2}))$ for $p=100$.}}
\vspace{0.2cm}
\begin{tabular}{r|r|rrrr|rrrr}
  \hline
  &&\multicolumn{4}{c|} {\bf learnDAG}&\multicolumn{4}{c} {\bf learnDAG.noPrun}\\
$p$&$n$& P & R&$F_1$ & time & P & R&$F_1$ & time   \\ 
    \hline
10&500    &   0.793 &   0.793 &   0.791 &   4.751 &   0.800 &   0.800 &   0.798 &   4.712 \\ 
  &1000 &   0.880 &   0.893 &   0.886 &   5.610 &   0.880 &   0.893 &   0.886 &   5.566 \\ 
  &2000 &   0.937 &   0.941 &   0.939 &   7.318 &   0.937 &   0.941 &   0.939 &   7.264 \\ 
  &&&&&&&&\\
100&500 &   0.878 &   0.659 &   0.753 & 43.458 &   0.878 &   0.662 &   0.755 & 43.213 \\ 
&1000&   0.914 &   0.792 &   0.848 & 69.872 &   0.913 &   0.795 &   0.850 & 69.630 \\ 
&2000 &   0.937 &   0.893 &   0.914 & 148.947 &   0.937 &   0.893 &   0.914 & 148.581 \\

  \hline
\end{tabular}
\end{table}

}

{\color{black}
In conclusion, it's crucial to note that algorithms such as learnDAG, which are based on modelling conditional distributions, overlook the consideration of marginal independences of variables. To illustrate the impact of potentially erroneously placing an edge between two marginally independent variables, a modified version of learnDAG, referred to as learnDAG.margin (incorporating marginal independence tests), could be considered. Specifically, an additional step, numbered 0 say, is introduced to initially test the marginal independence between all pairs of nodes, with the resulting graph serving as the input for step 1 in learnDAG. Some results are presented in Table \ref{table-margin}, reporting the Monte Carlo means of Precision (P), Recall (R), $F_1$-score, and running time of learnDAG and learnDAG.margin in some simulated scenarios. It is evident that at smaller sample sizes ($n=200$), the precision increases, indicating a higher probability of correctly identifying estimated edges. This is expected, as the possibility of erroneously placing an edge between two marginally independent variables is precluded. Conversely, for larger sample sizes ($n\ge 500$), benefits deriving from  a large sample size render step 0 irrelevant. Consequently, this finding implies that users could consider incorporating marginal independence tests when dealing with limited sample-sized data, while omitting this step for larger sample sizes.
}

\newpage
\appendix
\section{Figures and Tables}

\begin{figure}[http]
	\begin{center}
		\includegraphics[width = 0.9\linewidth, height=0.27\textheight]{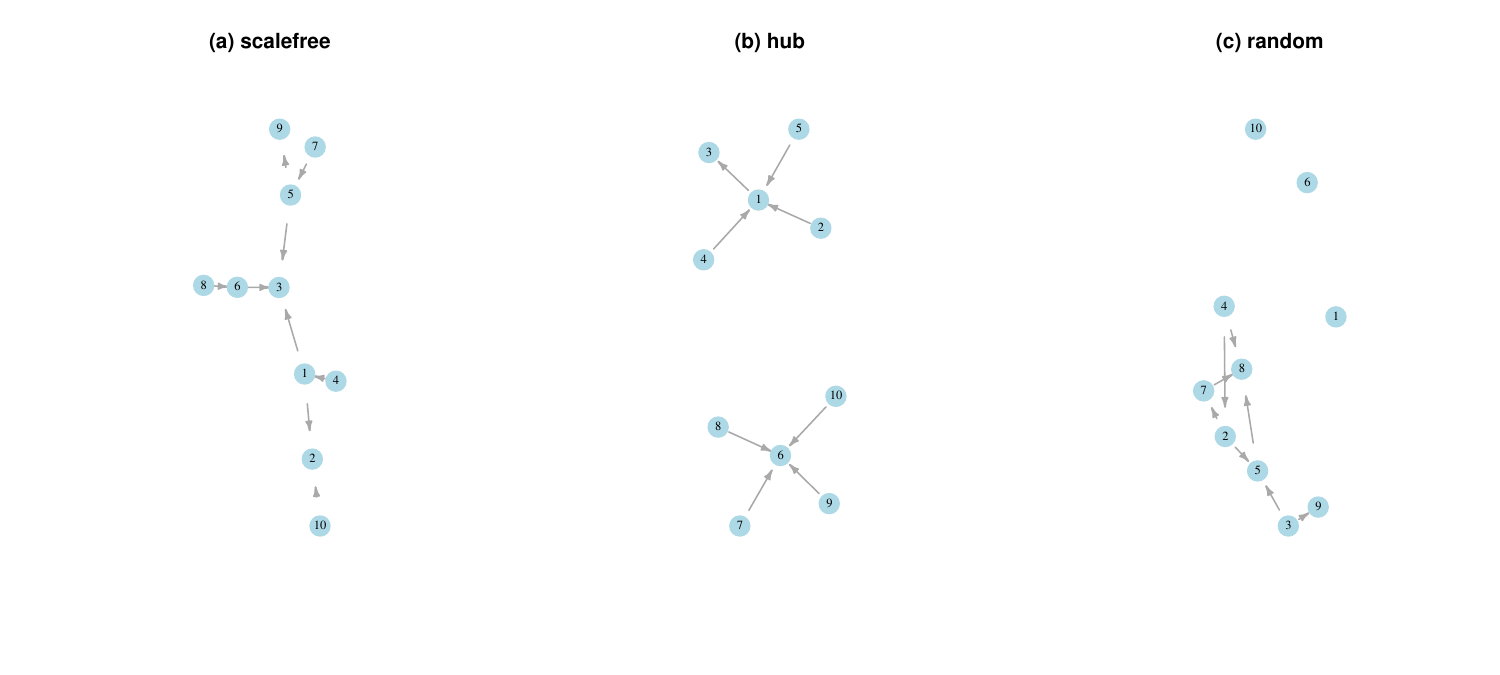}
		\vspace{-1cm}
		\caption{\scriptsize The graph structures for  $p=10$ employed in the simulation studies: (a) scale-free; (b)  hub; (c)  Erdos-Reny graph.}
		\label{DAGtypes10}
	\end{center}
\end{figure}

\begin{figure}[http]
	\begin{center}
		\includegraphics[width = 0.9\linewidth, height=0.25\textheight]{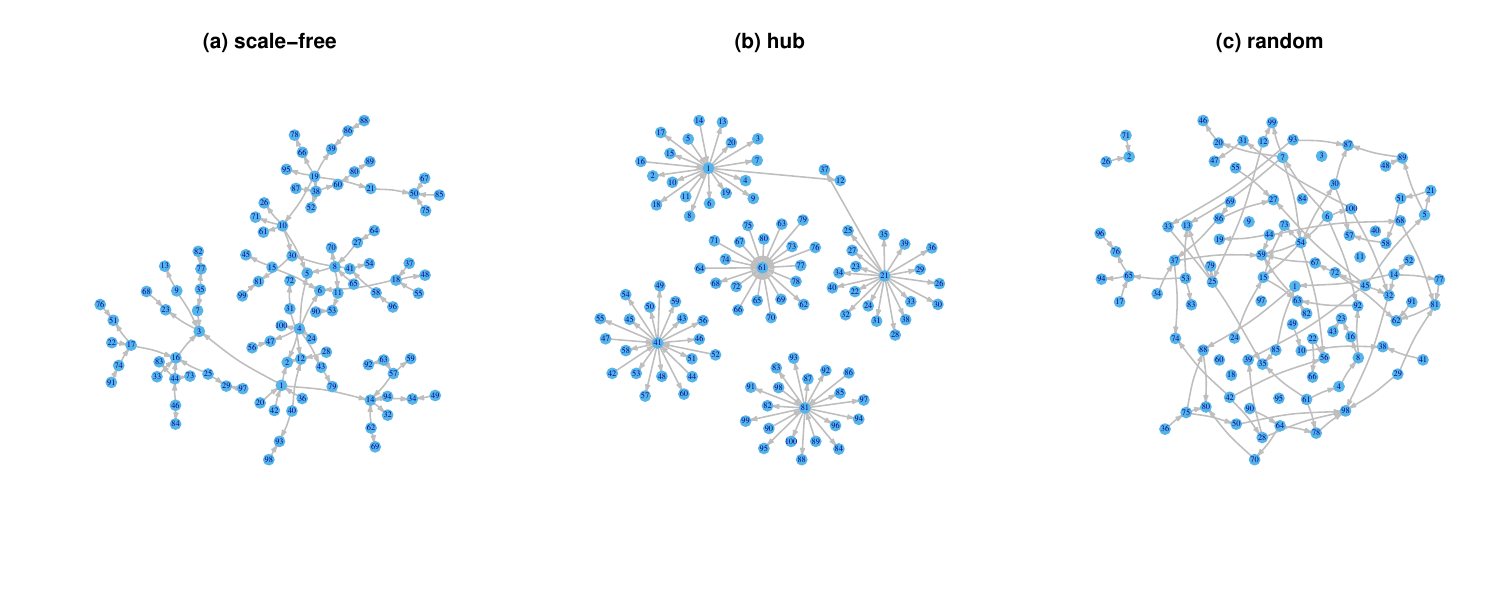}
		\vspace{-1cm}
		\caption{\scriptsize The graph structures for  $p=100$ employed in the simulation studies: (a) scale-free; (b)  hub; (c)  Erdos-Reny graph.}
		\label{DAGtypes100}
	\end{center}
\end{figure}
\begin{figure}[htbp]
	\begin{center}
		\includegraphics[width = 0.9\linewidth, height=0.43\textheight]{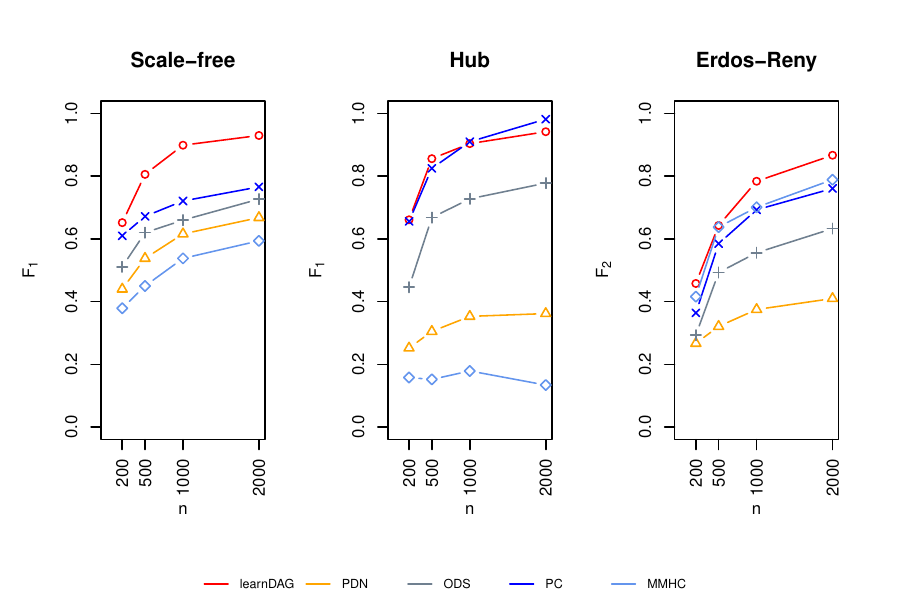}
		\caption{\scriptsize $F_1$-score of the considered algorithms: learnDAG; PDN; ODS; MMHC; PC; for the three types of graphs in Figure \ref{DAGtypes10} with $p=10$ and sample sizes $n=200,500,1000,2000.$}
		\label{DAG10F1}
	\end{center}
\end{figure}
\begin{figure}[htbp]
	\begin{center}
		\includegraphics[width = 0.9\linewidth, height=0.7\textheight]{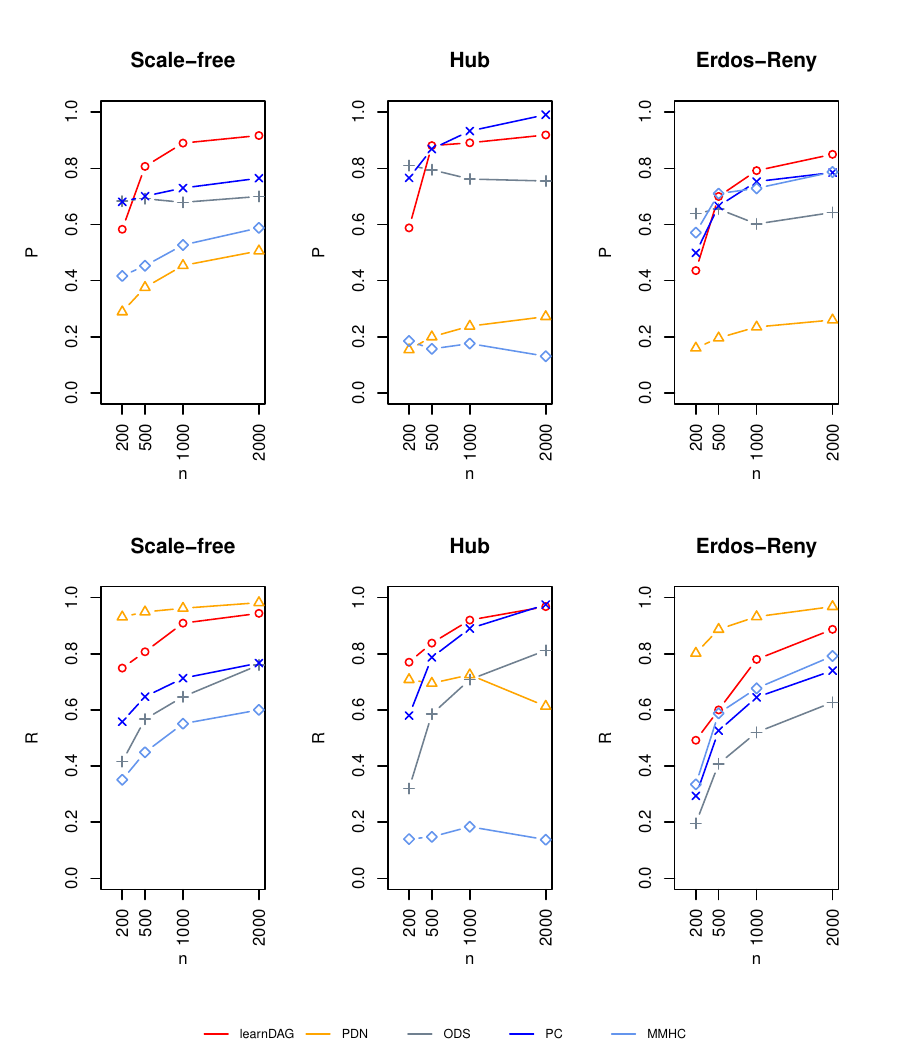}
		\caption{\scriptsize Precision and Recall of the considered algorithms: learnDAG; PDN; ODS; MMHC; PC; for the three types of graphs in Figure \ref{DAGtypes10} with $p=10$ and sample sizes $n=200,500,1000,2000.$}
		\label{DAG10PPVSe}
	\end{center}
\end{figure}

\newpage

\begin{center}
	\begin{scriptsize}
		\fontsize{6.2}{8}\selectfont
		\begin{longtable}{l| l | l r r r r r r r}
			\caption{Simulation results from 50 replicates of the DAGs shown in Figure \ref{DAGtypes10} in of the main paper  for $p= 10$ variables with Poisson node conditional distribution. Monte Carlo means (standard deviations) are shown for $TP, FP, FN, P, R$, and $F_1$. The levels of significance of tests $\alpha_{PNS}=\alpha_{Prun}=2(1-\Phi(n^{0.15}))$.}  \\
			\label{table10-chap2}\\
			\toprule
			Graph&$n$	& Algorithm & $TP$ & $FP$ & $FN$ & $P$ & $R$ &$F_1$&time \\
			\midrule
			\endfirsthead
			\multicolumn{9}{c}%
			{{\bfseries \tablename\ \thetable{} -- continued from previous page}} \\
			\toprule
			Graph&$n$	& Algorithm & $TP$ & $FP$ & $FN$ & $P$ & $R$ &$F_1$ &time \\
			\midrule	
			
			\endhead
  Scale-free &200&learnDAG & 6.740(1.322) & 5.020(2.035) & 2.260(1.322) & 0.583(0.135) & 0.749(0.147) & 0.652(0.133) &  9.595 \\ 
 & & PDN & 8.380(0.697) & 21.160(4.157) & 0.620(0.697) & 0.289(0.051) & 0.931(0.077) & 0.440(0.063) &  0.106 \\ 
 & & ODS & 3.740(0.922) & 1.880(1.206) & 5.260(0.922) & 0.683(0.181) & 0.416(0.102) & 0.511(0.118) &  3.841 \\ 
 & & PC & 5.020(0.892) & 2.400(1.010) & 3.980(0.892) & 0.681(0.103) & 0.558(0.099) & 0.610(0.093) &  0.007 \\ 
 & &  MMHC& 3.160(1.113) & 4.460(1.388) & 5.840(1.113) & 0.417(0.137) & 0.351(0.124) & 0.379(0.126) &  0.007 \\ 
 &&&& & & & & & \\
 &500&learnDAG& 7.260(1.139) & 1.760(1.170) & 1.740(1.139) & 0.807(0.124) & 0.807(0.127) & 0.806(0.121) &  9.396   \\
 & & PDN & 8.540(0.579) & 14.280(1.938) & 0.460(0.579) & 0.376(0.032) & 0.949(0.064) & 0.538(0.036) &  0.116 \\ 
 & & ODS & 5.100(1.035) & 2.360(1.258) & 3.900(1.035) & 0.692(0.142) & 0.567(0.115) & 0.620(0.118) &  4.717 \\ 
 & & PC & 5.820(0.850) & 2.500(0.863) & 3.180(0.850) & 0.701(0.096) & 0.647(0.094) & 0.672(0.092) &  0.008 \\ 
 & & MMHC & 4.040(1.340) & 4.820(1.224) & 4.960(1.340) & 0.453(0.137) & 0.449(0.149) & 0.450(0.142) &  0.008 \\ 
 &&&& & & & & & \\
 &1000 &learnDAG & 8.180(0.873) & 1.040(1.068) & 0.820(0.873) & 0.890(0.107) & 0.909(0.097) & 0.899(0.100) & 10.380 \\
 & & PDN & 8.660(0.479) & 10.480(1.359) & 0.340(0.479) & 0.454(0.034) & 0.962(0.053) & 0.616(0.036) &  0.137 \\ 
 & & ODS & 5.820(1.024) & 2.840(1.283) & 3.180(1.024) & 0.679(0.126) & 0.647(0.114) & 0.660(0.112) &  5.690 \\ 
 & & PC & 6.420(0.906) & 2.360(0.776) & 2.580(0.906) & 0.730(0.088) & 0.713(0.101) & 0.721(0.093) &  0.008 \\ 
 & & MMHC & 4.960(1.087) & 4.420(0.971) & 4.040(1.087) & 0.527(0.104) & 0.551(0.121) & 0.538(0.110) &  0.010 \\ 
  &&&& & & & & & \\
&2000 &learnDAG & 8.500(0.735) & 0.800(0.948) & 0.500(0.735) & 0.917(0.094) & 0.944(0.082) & 0.930(0.085) & 10.884 \\ 
 & & PDN & 8.840(0.370) & 8.680(1.096) & 0.160(0.370) & 0.506(0.033) & 0.982(0.041) & 0.668(0.033) &  0.176 \\ 
 & & ODS& 6.840(1.095) & 2.980(1.270) & 2.160(1.095) & 0.700(0.120) & 0.760(0.122) & 0.727(0.116) &  7.668 \\ 
 & & PC & 6.900(0.364) & 2.120(0.385) & 2.100(0.364) & 0.765(0.042) & 0.767(0.040) & 0.766(0.041) &  0.008 \\ 
 & & MMHC  & 5.400(0.881) & 3.800(0.969) & 3.600(0.881) & 0.588(0.099) & 0.600(0.098) & 0.594(0.098) &  0.012 \\
 \hline
 &&&& & & & & & \\
	Hub &200&learnDAG& 6.160(1.184) & 4.580(1.970) & 1.840(1.184) & 0.588(0.140) & 0.770(0.148) & 0.661(0.130) &  9.486\\
  & & PDN & 5.660(1.171) & 31.240(3.679) & 2.340(1.171) & 0.154(0.030) & 0.708(0.146) & 0.252(0.049) &  0.107 \\ 
  & & ODS & 2.560(0.951) & 0.700(0.789) & 5.440(0.951) & 0.810(0.193) & 0.320(0.119) & 0.446(0.136) &  4.002 \\ 
  & & PC & 4.640(1.102) & 1.400(0.808) & 3.360(1.102) & 0.766(0.138) & 0.580(0.138) & 0.656(0.133) &  0.008 \\ 
  & &MMHC & 1.122(0.726) & 5.082(1.272) & 6.878(0.726) & 0.185(0.105) & 0.140(0.091) & 0.158(0.096) &  0.007 \\ 
   &&&& & & & & & \\
  &500 & learnDAG & 6.700(0.863) & 0.940(0.867) & 1.300(0.863) & 0.881(0.106) & 0.838(0.108) & 0.856(0.096) &  9.505\\
  & & PDN & 5.560(1.198) & 22.960(4.755) & 2.440(1.198) & 0.200(0.045) & 0.695(0.150) & 0.305(0.053) &  0.118 \\ 
  & & ODS & 4.680(1.096) & 1.280(1.089) & 3.320(1.096) & 0.795(0.164) & 0.585(0.137) & 0.668(0.136) &  5.239 \\ 
  & & PC & 6.300(1.129) & 1.000(1.294) & 1.700(1.129) & 0.869(0.162) & 0.787(0.141) & 0.825(0.147) &  0.009 \\ 
  & & MMHC & 1.188(0.641) & 6.354(1.041) & 6.812(0.641) & 0.157(0.072) & 0.148(0.080) & 0.152(0.075) &  0.008 \\ 
   &&&& & & & & & \\
  &1000&learnDAG & 7.360(0.631) & 0.940(0.818) & 0.640(0.631) & 0.891(0.088) & 0.920(0.079) & 0.904(0.076) & 10.413\\
  & & PDN & 5.800(1.370) & 18.880(3.910) & 2.200(1.370) & 0.238(0.044) & 0.725(0.171) & 0.353(0.065) &  0.138 \\ 
  & & ODS & 5.660(0.872) & 1.940(1.346) & 2.340(0.872) & 0.762(0.140) & 0.708(0.109) & 0.728(0.104) &  7.062 \\ 
  & & PC & 7.120(1.003) & 0.540(1.110) & 0.880(1.003) & 0.933(0.131) & 0.890(0.125) & 0.910(0.125) &  0.009 \\ 
  & &MMHC & 1.468(1.080) & 6.660(1.069) & 6.532(1.080) & 0.176(0.113) & 0.184(0.135) & 0.179(0.123) &  0.009\\
     &&&& & & & & & \\
   &2000&learnDAG& 7.740(0.443) & 0.720(0.730) & 0.260(0.443) & 0.919(0.079) & 0.968(0.055) & 0.942(0.060) & 10.872 \\   
  & & PDN  & 4.900(1.502) & 13.500(4.253) & 3.100(1.502) & 0.272(0.053) & 0.613(0.188) & 0.362(0.073) &  0.177 \\ 
  & & ODS  & 6.500(0.789) & 2.260(1.226) & 1.500(0.789) & 0.755(0.123) & 0.812(0.099) & 0.778(0.091) & 10.706 \\ 
  & & PC & 7.800(0.404) & 0.080(0.340) & 0.200(0.404) & 0.991(0.038) & 0.975(0.051) & 0.982(0.036) &  0.010 \\ 
  & &MMHC  & 1.102(0.510) & 7.286(0.816) & 6.898(0.510) & 0.131(0.053) & 0.138(0.064) & 0.134(0.058) &  0.011 \\ 
  \hline
   &&&& & & & & & \\
 Erdos-Reny  &200&learnDAG  &3.940(1.361) & 5.320(2.065) & 4.060(1.361) & 0.436(0.158) & 0.492(0.170) & 0.458(0.156) &  9.541 \\
  & & PDN & 6.420(1.012) & 34.060(4.688) & 1.580(1.012) & 0.160(0.029) & 0.802(0.126) & 0.267(0.045) &  0.108 \\ 
  & & ODS & 1.568(0.695) & 1.114(0.993) & 6.432(0.695) & 0.639(0.280) & 0.196(0.087) & 0.293(0.120) &  3.919 \\ 
 & & PC & 2.356(1.282) & 2.222(0.974) & 5.644(1.282) & 0.499(0.193) & 0.294(0.160) & 0.364(0.174) &  0.006\\ 
  & & MMHC & 2.680(1.115) & 2.060(1.185) & 5.320(1.115) & 0.571(0.191) & 0.335(0.139) & 0.416(0.155) &  0.006 \\ 
   &&&& & & & & & \\
  &500 &learnDAG & 4.800(1.355) & 2.080(1.291) & 3.200(1.355) & 0.700(0.183) & 0.600(0.169) & 0.643(0.169) &  9.454 \\ 
  & & PDN & 7.100(0.735) & 29.300(3.066) & 0.900(0.735) & 0.196(0.025) & 0.887(0.092) & 0.321(0.039) &  0.119 \\ 
 & &  ODS & 3.260(1.275) & 1.800(1.262) & 4.740(1.275) & 0.655(0.209) & 0.408(0.159) & 0.493(0.168) &  4.924 \\ 
  & & PC & 4.204(1.399) & 2.020(0.989) & 3.796(1.399) & 0.666(0.174) & 0.526(0.175) & 0.585(0.173) &  0.007  \\ 
  & & MMHC & 4.700(1.055) & 2.000(1.178) & 3.300(1.055) & 0.710(0.140) & 0.588(0.132) & 0.638(0.126) &  0.007 \\ 
   &&&& & & & & & \\
  &1000&learnDAG  & 6.240(1.222) & 1.680(1.253) & 1.760(1.222) & 0.792(0.151) & 0.780(0.153) & 0.784(0.145) & 10.331\\
  & & PDN & 7.460(0.676) & 24.420(2.417) & 0.540(0.676) & 0.235(0.024) & 0.932(0.085) & 0.375(0.035) &  0.139 \\ 
  & & ODS & 4.160(1.315) & 2.800(1.414) & 3.840(1.315) & 0.602(0.186) & 0.520(0.164) & 0.556(0.169) &  6.144 \\ 
  & & PC & 5.160(1.095) & 1.640(0.631) & 2.840(1.095) & 0.753(0.111) & 0.645(0.137) & 0.693(0.124) &  0.008 \\ 
  & & MMHC & 5.420(0.992) & 2.040(1.029) & 2.580(0.992) & 0.729(0.129) & 0.677(0.124) & 0.701(0.121) &  0.009 \\ 

       &&&& & & & & & \\
  &2000&learnDAG & 7.100(1.111) & 1.280(1.161) & 0.900(1.111) & 0.850(0.137) & 0.887(0.139) & 0.867(0.133) & 10.799 \\
  & & PDN & 7.740(0.487) & 22.120(2.086) & 0.260(0.487) & 0.260(0.023) & 0.968(0.061) & 0.410(0.032) &  0.179 \\ 
  & & ODS& 5.020(1.532) & 2.900(1.821) & 2.980(1.532) & 0.643(0.203) & 0.627(0.191) & 0.633(0.192) &  8.243 \\ 
  & & PC & 5.920(0.922) & 1.600(0.670) & 2.080(0.922) & 0.785(0.095) & 0.740(0.115) & 0.761(0.103) &  0.008 \\ 
  & & MMHC  & 6.340(0.717) & 1.740(0.803) & 1.660(0.717) & 0.787(0.090) & 0.792(0.090) & 0.789(0.086) &  0.012 \\ 
			\bottomrule
		\end{longtable}
	\end{scriptsize}
	
\end{center}

\begin{center}
	\begin{scriptsize}
	\fontsize{6.2}{6}\selectfont
		\begin{longtable}{l| l | l r r r r r r r}
			\caption{Simulation results from 50 replicates of the DAGs shown in Figure \ref{DAGtypes100} in of the main paper  for $p= 100$ variables with Poisson node conditional distribution. Monte Carlo means (standard deviations) are shown for $TP, FP, FN, P, R$, and $F_1$. The levels of significance of tests $\alpha_{PNS}=\alpha_{Prun}=2(1-\Phi(n^{0.2}))))$.}  \\
			\label{table100-chap2}\\
			\toprule
			Graph&$n$	& Algorithm & $TP$ & $FP$ & $FN$ & $P$ & $R$ &$F_1$&time \\
			\midrule
			\endfirsthead
			\multicolumn{9}{c}%
			{{\bfseries \tablename\ \thetable{} -- continued from previous page}} \\
			\toprule
			Graph&$n$	& Algorithm & $TP$ & $FP$ & $FN$ & $P$ & $R$ &$F_1$& time \\
			\midrule	
			
			\endhead
  Scale-free&500&learnDAG& 65.600(2.390) & 8.860(3.031) & 33.400(2.390) & 0.882(0.038) & 0.663(0.024) & 0.756(0.027) &  45.888 \\ 
  &&PDN & 48.520(23.784) & 111.880(87.858) & 50.480(23.784) & 0.417(0.202) & 0.490(0.240) & 0.351(0.134) &   4.831 \\ 
  &&ODS & 45.740(5.938) & 94.840(32.950) & 53.260(5.938) & 0.340(0.062) & 0.462(0.060) & 0.385(0.039) & 195.076 \\ 
  &&PC & 39.940(3.158) & 28.800(3.090) & 59.060(3.158) & 0.581(0.040) & 0.403(0.032) & 0.476(0.034) &   0.123 \\ 
  &&MMHC & 55.220(5.152) & 86.880(7.634) & 43.780(5.152) & 0.389(0.037) & 0.558(0.052) & 0.458(0.042) &   0.530 \\ 
  &&&& & & & & & \\
  &1000&learnDAG& 82.500(3.138) & 11.180(3.205) & 16.500(3.138) & 0.881(0.033) & 0.833(0.032) & 0.856(0.029) &  76.931 \\
  &&PDN & 45.980(26.072) & 63.160(52.492) & 53.020(26.072) & 0.538(0.197) & 0.464(0.263) & 0.399(0.162) &   6.473 \\ 
  &&ODS & 51.580(12.993) & 84.300(31.210) & 47.420(12.993) & 0.404(0.090) & 0.521(0.131) & 0.431(0.083) & 307.384 \\ 
  &&PC & 46.880(2.862) & 32.740(2.841) & 52.120(2.862) & 0.589(0.031) & 0.474(0.029) & 0.525(0.028) &   0.152 \\ 
  &&MMHC & 66.760(4.770) & 73.200(7.157) & 32.240(4.770) & 0.478(0.040) & 0.674(0.048) & 0.559(0.043) &   0.825 \\ 
  &&&& & & & & & \\
  &2000&learnDAG& 90.400(2.634) & 10.140(2.857) & 8.600(2.634) & 0.899(0.027) & 0.913(0.027) & 0.906(0.026) & 145.481\\
  &&PDN & 34.540(23.652) & 33.940(28.183) & 64.460(23.652) & 0.594(0.160) & 0.349(0.239) & 0.360(0.189) &  10.075 \\ 
  &&ODS & 61.840(13.864) & 95.280(32.128) & 37.160(13.864) & 0.403(0.074) & 0.625(0.140) & 0.477(0.088) & 516.160 \\ 
  &&PC & 54.200(2.563) & 35.640(2.855) & 44.800(2.563) & 0.603(0.029) & 0.547(0.026) & 0.574(0.027) &   0.195 \\ 
  &&MMHC & 73.900(4.082) & 60.580(7.535) & 25.100(4.082) & 0.551(0.042) & 0.746(0.041) & 0.634(0.041) &   1.396 \\ 
  &&&& & & & & & \\
   &5000&learnDAG& 93.940(2.014) & 8.560(2.620) & 5.060(2.014) & 0.917(0.025) & 0.949(0.020) & 0.932(0.021) & 380.035\\
 &&PDN  & 22.771(18.227) & 19.771(19.630) & 76.229(18.227) & 0.646(0.187) & 0.230(0.184) & 0.277(0.185) &   21.215  \\ 
  &&ODS  & 65.820(17.052) & 78.120(31.131) & 33.180(17.052) & 0.478(0.101) & 0.665(0.172) & 0.532(0.102) &  938.877 \\ 
 &&PC & 59.960(1.355) & 37.640(1.903) & 39.040(1.355) & 0.614(0.017) & 0.606(0.014) & 0.610(0.015) &    0.250 \\ 
 &&MMHC & 76.390(3.420) & 51.341(4.768) & 22.610(3.420) & 0.598(0.031) & 0.772(0.035) & 0.674(0.031) &    3.690 \\
\hline
&&&& & & & & & \\
 Hub&500&learnDAG&27.551(4.435) &10.510(3.117) & 67.449(4.435) & 0.723(0.078) & 0.290(0.047) & 0.413(0.057) &47.281\\
 && PDN & 60.980(3.396) & 653.320(42.419) & 34.020(3.396) & 0.086(0.005) & 0.642(0.036) & 0.151(0.008) &    4.852 \\ 
 && ODS & 18.340(8.178) & 72.520(28.367) & 76.660(8.178) & 0.211(0.101) & 0.193(0.086) & 0.194(0.079) &  145.684 \\ 
 && PC & 5.000(1.539) & 24.060(2.189) & 90.000(1.539) & 0.171(0.049) & 0.053(0.016) & 0.080(0.024) &    0.350 \\ 
 && MMHC & 37.260(4.580) & 84.720(10.236) & 57.740(4.580) & 0.307(0.046) & 0.392(0.048) & 0.344(0.046) &    9.120 \\ 
  &&&& & & & & & \\
 &1000& learnDAG&53.120(3.874) & 21.100(3.786) & 41.880(3.874) & 0.716(0.046) & 0.559(0.041) & 0.628(0.041) &  76.040\\
 && PDN & 69.200(2.441) & 546.600(20.849) & 25.800(2.441) & 0.112(0.005) & 0.728(0.026) & 0.195(0.008) &    6.503 \\ 
 && ODS & 26.280(9.311) & 92.760(12.697) & 68.720(9.311) & 0.221(0.076) & 0.277(0.098) & 0.245(0.085) &  232.417 \\ 
 && PC & 7.760(1.533) & 34.260(2.221) & 87.240(1.533) & 0.185(0.036) & 0.082(0.016) & 0.113(0.022) &    4.832 \\ 
 && MMHC & 53.740(4.931) & 71.900(9.429) & 41.260(4.931) & 0.429(0.047) & 0.566(0.052) & 0.488(0.047) &  358.392 \\ 
  &&&& & & & & & \\
 &2000& learnDAG &68.480(3.587) & 24.480(3.059) & 26.520(3.587) & 0.737(0.032) & 0.721(0.038) & 0.729(0.034) & 139.437 \\  
 && PDN & 71.100(2.188) & 421.220(30.434) & 23.900(2.188) & 0.145(0.009) & 0.748(0.023) & 0.243(0.013) &   10.005 \\ 
 && ODS & 36.020(11.302) & 82.320(12.173) & 58.980(11.302) & 0.304(0.093) & 0.379(0.119) & 0.337(0.104) &  397.424 \\ 
 && PC & 12.460(2.111) & 46.800(2.304) & 82.540(2.111) & 0.210(0.029) & 0.131(0.022) & 0.161(0.025) &   86.424 \\ 
 && MMHC & 66.000(7.323) & 57.440(10.459) & 29.000(7.323) & 0.537(0.069) & 0.695(0.077) & 0.605(0.072) & 3351.923 \\ 
 &&&& & & & & & \\
 &5000& learnDAG&75.480(3.105) & 24.400(3.653) & 19.520(3.105) & 0.756(0.034) & 0.795(0.033) & 0.775(0.032) & 356.568 \\
 && PDN  & 71.620(1.576) & 239.500(26.011) & 23.380(1.576) & 0.232(0.024) & 0.754(0.017) & 0.354(0.027) &    20.979 \\ 
 && ODS & 46.260(11.194) & 55.680(11.757) & 48.740(11.194) & 0.454(0.112) & 0.487(0.118) & 0.470(0.114) &   790.223\\ 
 && PC & 21.000(1.678) & 60.120(1.452) & 74.000(1.678) & 0.259(0.015) & 0.221(0.018) & 0.238(0.017) &  1031.772 \\ 
 && MMHC & 70.682(7.414) & 46.045(8.518) & 24.318(7.414) & 0.606(0.066) & 0.744(0.078) & 0.668(0.071) & 12170.984 \\ 
 \hline
 &&&& & & & & & \\
Erdos-Reny&500& learnDAG& 51.960(4.150) & 15.860(3.540) & 57.040(4.150) & 0.766(0.050) & 0.477(0.038) & 0.587(0.041) &  45.768  \\
 &&  PDN & 84.820(3.652) & 494.120(20.495) & 24.180(3.652) & 0.147(0.008) & 0.778(0.034) & 0.247(0.013) &   4.825 \\ 
  && ODS & 44.040(4.458) & 84.660(10.809) & 64.960(4.458) & 0.344(0.038) & 0.404(0.041) & 0.371(0.036) & 158.765 \\ 
 &&  PC & 35.140(3.175) & 25.520(2.816) & 73.860(3.175) & 0.579(0.039) & 0.322(0.029) & 0.414(0.032) &   0.100 \\ 
 &&  MMHC & 51.500(5.632) & 96.420(9.476) & 57.500(5.632) & 0.349(0.040) & 0.472(0.052) & 0.401(0.043) &   0.576 \\
  &&&& & & & & & \\
&1000&learnDAG&77.900(4.748) & 19.840(4.510) & 31.100(4.748) & 0.797(0.045) & 0.715(0.044) & 0.753(0.043) &  75.111 \\
 &&  PDN & 93.960(2.725) & 361.620(18.856) & 15.040(2.725) & 0.207(0.011) & 0.862(0.025) & 0.333(0.015) &   6.445 \\ 
 &&  ODS & 58.380(5.810) & 74.780(8.115) & 50.620(5.810) & 0.439(0.045) & 0.536(0.053) & 0.482(0.048) & 243.041 \\ 
 &&  PC & 47.620(2.996) & 31.480(3.032) & 61.380(2.996) & 0.602(0.034) & 0.437(0.027) & 0.506(0.029) &   0.117 \\
&&  MMHC & 66.840(4.892) & 79.260(7.537) & 42.160(4.892) & 0.458(0.037) & 0.613(0.045) & 0.524(0.039) &   0.943 \\ 
  &&&& & & & & & \\
 &2000&learnDAG& 89.800(3.090) & 18.720(3.214) & 19.200(3.090) & 0.828(0.028) & 0.824(0.028) & 0.826(0.027) & 139.479\\
  && PDN & 97.560(1.864) & 257.700(11.438) & 11.440(1.864) & 0.275(0.010) & 0.895(0.017) & 0.421(0.012) &   9.936 \\ 
 &&  ODS & 66.440(4.625) & 70.460(7.276) & 42.560(4.625) & 0.486(0.039) & 0.610(0.042) & 0.541(0.040) & 385.807 \\ 
 &&  PC & 57.540(3.303) & 35.100(2.150) & 51.460(3.303) & 0.621(0.025) & 0.528(0.030) & 0.571(0.027) &   0.141 \\ 
 &&  MMHC & 76.400(4.412) & 64.420(7.680) & 32.600(4.412) & 0.544(0.040) & 0.701(0.040) & 0.612(0.039) &   1.797 \\ 
   &&&& & & & & & \\
 &5000&learnDAG& 97.260(2.586) & 14.300(3.209) & 11.740(2.586) & 0.872(0.028) & 0.892(0.024) & 0.882(0.025) & 356.971\\
  && PDN & 98.980(2.075) & 183.600(8.069) & 10.020(2.075) & 0.351(0.011) & 0.908(0.019) & 0.506(0.013) &  20.875 \\ 
 &&  ODS & 77.380(5.174) & 68.760(31.428) & 31.620(5.174) & 0.551(0.109) & 0.710(0.047) & 0.615(0.081) & 779.215 \\ 
 &&  PC & 65.860(2.080) & 38.920(1.652) & 43.140(2.080) & 0.629(0.016) & 0.604(0.019) & 0.616(0.017) &   0.177 \\ 
 &&  MMHC & 85.350(3.371) & 44.125(6.014) & 23.650(3.371) & 0.660(0.037) & 0.783(0.031) & 0.716(0.033) &   3.712 \\ 
			
			\bottomrule
			
		\end{longtable}
	\end{scriptsize}
	
\end{center}

 \bibliographystyle{elsarticle-num} 
 \bibliography{main}





\end{document}